\documentclass[ALICE,manyauthors]{cernphprep}
\usepackage[comma,square,numbers,sort&compress]{natbib}

\usepackage{lineno}
\usepackage{xspace}
\usepackage{hyperref}
\usepackage[T1]{fontenc}
\usepackage{orcidlink}

\begin{document}
%

\newcommand{\pp}           {pp\xspace}
\newcommand{\ppbar}        {\mbox{$\mathrm {p\overline{p}}$}\xspace}
\newcommand{\XeXe}         {\mbox{Xe--Xe}\xspace}
\newcommand{\PbPb}         {\mbox{Pb--Pb}\xspace}
\newcommand{\pA}           {\mbox{pA}\xspace}
\newcommand{\pPb}          {\mbox{p--Pb}\xspace}
\newcommand{\AuAu}         {\mbox{Au--Au}\xspace}
\newcommand{\dAu}          {\mbox{d--Au}\xspace}

\newcommand{\mee}         {\ensuremath{m_{\rm ee}}\xspace}
\newcommand{\ptee}         {\ensuremath{p_{\rm T,ee}}\xspace}
\newcommand{\pteesquare}    {\ensuremath{p_{\rm T,ee}^{\rm 2}}\xspace}
\newcommand{\meanpteesquare}       {$\langle p_{\rm T,ee}^{\rm 2}\rangle$\xspace}
\newcommand{\s}            {\ensuremath{\sqrt{s}}\xspace}
\newcommand{\snn}          {\ensuremath{\sqrt{s_{\mathrm{NN}}}}\xspace}
\newcommand{\pt}           {\ensuremath{p_{\rm T}}\xspace}
\newcommand{\pte}           {\ensuremath{p_{\rm T,e}}\xspace}
\newcommand{\meanpt}       {$\langle p_{\mathrm{T}}\rangle$\xspace}
\newcommand{\ycms}         {\ensuremath{y_{\rm CMS}}\xspace}
\newcommand{\ylab}         {\ensuremath{y_{\rm lab}}\xspace}
\newcommand{\etarange}[1]  {\mbox{$| \eta |\,<\,#1$}}
\newcommand{\etarangee}[1]  {\mbox{$| \eta_{\rm e} |<#1$}}
\newcommand{\yrange}[1]    {\mbox{$\left | y \right |~<~#1$}}
\newcommand{\dndy}         {\ensuremath{\mathrm{d}N_\mathrm{ch}/\mathrm{d}y}\xspace}
\newcommand{\dndeta}       {\ensuremath{\mathrm{d}N_\mathrm{ch}/\mathrm{d}\eta}\xspace}
\newcommand{\avdndeta}     {\ensuremath{\langle\dndeta\rangle}\xspace}
\newcommand{\dNdy}         {\ensuremath{\mathrm{d}N_\mathrm{ch}/\mathrm{d}y}\xspace}
\newcommand{\Npart}        {\ensuremath{N_\mathrm{part}}\xspace}
\newcommand{\Ncoll}        {\ensuremath{N_\mathrm{coll}}\xspace}
\newcommand{\dEdx}         {\ensuremath{\textrm{d}E/\textrm{d}x}\xspace}
\newcommand{\RpPb}         {\ensuremath{R_{\rm pPb}}\xspace}

\newcommand{\nineH}        {$\sqrt{s}~=~0.9$~Te\kern-.1emV\xspace}
\newcommand{\seven}        {$\sqrt{s}~=~7$~Te\kern-.1emV\xspace}
\newcommand{\twoH}         {$\sqrt{s}~=~0.2$~Te\kern-.1emV\xspace}
\newcommand{\twosevensix}  {$\sqrt{s}~=~2.76$~Te\kern-.1emV\xspace}
\newcommand{\five}         {$\sqrt{s}~=~5.02$~Te\kern-.1emV\xspace}
\newcommand{\twosevensixnn}{$\sqrt{s_{\mathrm{NN}}}~=~2.76$~Te\kern-.1emV\xspace}
\newcommand{\fivenn}       {$\sqrt{s_{\mathrm{NN}}}~=~5.02$~Te\kern-.1emV\xspace}
\newcommand{\LT}           {L{\'e}vy-Tsallis\xspace}
\newcommand{\GeVc}         {Ge\kern-.1emV/$c$\xspace}
\newcommand{\MeVc}         {Me\kern-.1emV/$c$\xspace}
\newcommand{\TeV}          {Te\kern-.1emV\xspace}
\newcommand{\GeV}          {Ge\kern-.1emV\xspace}
\newcommand{\MeV}          {Me\kern-.1emV\xspace}
\newcommand{\GeVmass}      {Ge\kern-.1emV/$c^2$\xspace}
\newcommand{\MeVmass}      {Me\kern-.2emV/$c^2$\xspace}
\newcommand{\lumi}         {\ensuremath{\mathcal{L}}\xspace}

\newcommand{\ITS}          {\rm{ITS}\xspace}
\newcommand{\TOF}          {\rm{TOF}\xspace}
\newcommand{\ZDC}          {\rm{ZDC}\xspace}
\newcommand{\ZDCs}         {\rm{ZDCs}\xspace}
\newcommand{\ZNA}          {\rm{ZNA}\xspace}
\newcommand{\ZNC}          {\rm{ZNC}\xspace}
\newcommand{\SPD}          {\rm{SPD}\xspace}
\newcommand{\SDD}          {\rm{SDD}\xspace}
\newcommand{\SSD}          {\rm{SSD}\xspace}
\newcommand{\TPC}          {\rm{TPC}\xspace}
\newcommand{\TRD}          {\rm{TRD}\xspace}
\newcommand{\VZERO}        {\rm{V0}\xspace}
\newcommand{\VZEROA}       {\rm{V0A}\xspace}
\newcommand{\VZEROC}       {\rm{V0C}\xspace}
\newcommand{\Vdecay} 	   {\ensuremath{V^{0}}\xspace}

\newcommand{\ee}           {\ensuremath{{\rm e}^{+}{\rm e}^{-}}\xspace} 
\newcommand{\pip}          {\ensuremath{\pi^{+}}\xspace}
\newcommand{\pim}          {\ensuremath{\pi^{-}}\xspace}
\newcommand{\kap}          {\ensuremath{\rm{K}^{+}}\xspace}
\newcommand{\kam}          {\ensuremath{\rm{K}^{-}}\xspace}
\newcommand{\pbar}         {\ensuremath{\rm\overline{p}}\xspace}
\newcommand{\kzero}        {\ensuremath{{\rm K}^{0}_{\rm{S}}}\xspace}
\newcommand{\lmb}          {\ensuremath{\Lambda}\xspace}
\newcommand{\almb}         {\ensuremath{\overline{\Lambda}}\xspace}
\newcommand{\Om}           {\ensuremath{\Omega^-}\xspace}
\newcommand{\Mo}           {\ensuremath{\overline{\Omega}^+}\xspace}
\newcommand{\X}            {\ensuremath{\Xi^-}\xspace}
\newcommand{\Ix}           {\ensuremath{\overline{\Xi}^+}\xspace}
\newcommand{\Xis}          {\ensuremath{\Xi^{\pm}}\xspace}
\newcommand{\Oms}          {\ensuremath{\Omega^{\pm}}\xspace}
\newcommand{\degree}       {\ensuremath{^{\rm o}}\xspace}

\begin{titlepage}
\PHyear{2022}       
\PHnumber{068}      
\PHdate{24 March}  

\title{Dielectron production at midrapidity at low transverse momentum in peripheral and semi-peripheral Pb--Pb collisions at {\bf{\fivenn}}}
\ShortTitle{Low \ptee dielectrons in peripheral Pb--Pb collisions at \fivenn}   

\Collaboration{ALICE Collaboration\thanks{See Appendix~\ref{app:collab} for the list of collaboration members}}
\ShortAuthor{ALICE Collaboration} 

\begin{abstract}
The first measurement of the \ee pair production at low lepton pair transverse momentum (\ptee) and low invariant mass (\mee) in non-central \PbPb collisions at \fivenn at the LHC is presented. The dielectron production is studied with the ALICE detector at midrapidity (\etarangee{0.8}) as a function of invariant mass (0.4 $\leq$ \mee $<$ 2.7~\GeVmass) in the 50--70\% and 70--90\% centrality classes for \ptee $<$ 0.1~\GeVc, and as a function of \ptee in three \mee intervals in the most peripheral \PbPb collisions. Below a \ptee of 0.1~\GeVc, a clear excess of \ee pairs is found compared to the expectations from known hadronic sources and predictions of thermal radiation from the medium. The \mee excess spectra are reproduced, within uncertainties, by different predictions of the photon--photon production of dielectrons, where the photons originate from the extremely strong electromagnetic fields generated by the highly Lorentz-contracted Pb nuclei. Lowest-order quantum electrodynamic (QED) calculations, as well as a model that takes into account the impact-parameter dependence of the average transverse momentum of the photons, also provide a good description of the \ptee spectra. The measured $\sqrt{\langle \ptee^{2} \rangle}$ of the excess \ptee spectrum in peripheral \PbPb collisions is found to be comparable to the values observed previously at RHIC in a similar phase-space region.

\end{abstract}
\end{titlepage}

\setcounter{page}{2} 


\section{Introduction} 

Ultra-relativistic heavy-ion collisions produce the largest electromagnetic (EM) fields experimentally accessible in the universe. The magnetic field generated by the highly Lorentz-contracted passing nuclei is predicted to reach up to 10$^{\rm 15}$~Tesla\,\cite{Kharzeev:2007jp}. Such strong EM fields are predicted to produce various exotic phenomena\,\cite{Asakawa:2010bu,Hattori:2012je,Koch:2016pzl,Battesti:2018bgc}. Heavy-ion collisions have therefore, in the past decades, induced a large amount of experimental and theoretical interest in the search for new aspects of quantum chromodynamics (QCD) and quantum electrodynamics (QED)\,\cite{Kharzeev:2015znc,STAR:2009wot,ALICE:2012nhw,CMS:2016wfo}.

The measurement of thermal dileptons from the quark--gluon plasma and the hot hadron gas produced in heavy-ion collisions has been long recognized as a clean and powerful probe to study the time evolution of the properties of the medium. Another important dilepton production mechanism, in particular at very low lepton pair transverse momentum ($p_{\rm T,ll}$), is the photon--photon fusion process ($\gamma\gamma \to {\rm l^{+}} {\rm l^{-}}$). The EM fields surrounding the relativistic heavy ions with large charge number $Z$ can be treated as a flux of quasi-real photons generated coherently, i.e.\ the charges of the $Z$ protons in the nucleus act coherently leading to a $Z^{\rm 2}$ dependence of the quasi-real photon flux. Such photons, triggered by the EM fields of the two incoming nuclei, can interact via the Breit--Wheeler process\,\cite{Breit:1934zz} to produce dileptons. Such an exclusive photon-mediated process was first measured in ultra-peripheral heavy-ion collisions (UPC) by the STAR collaboration at RHIC\,\cite{STAR:2004bzo}. Collisions with impact parameters ($b$) between the passing nuclei large enough that no nuclear overlap occurs can be selected, excluding any hadronic interaction. Only recently, the photon--photon production of dileptons has been observed in hadronic heavy-ion collisions (HHIC) by the STAR\,\cite{STAR:2018ldd} and ATLAS\,\cite{ATLAS:2018pfw,ATLAS:2022vbe} collaborations. STAR measures dielectrons ($\rm e^{+}\rm e^{-}$) at midrapidity and small invariant mass \mee (0.4 $\leq$ \mee $\leq$ 2.6~\GeVmass) in non-central Au--Au and U--U collisions at a center-of-mass energy per nucleon pair of $\sqrt{s_{\rm NN}}$ $=$ 200~GeV and 193~GeV, respectively, whereas ATLAS reports results on dimuon ($\mu^{+}\mu^{-}$) production at large $m_{\rm \mu\mu}$ ( 4 $\leq$ $m_{\rm \mu\mu}$ $<$ 45~\GeVmass) in central, semi-central and peripheral \PbPb collisions at \fivenn. The produced dileptons originate from quasi-real photons with momenta predominantly in the beam direction, i.e.\ the transverse component is of the order of $\omega_{\gamma}$/$\gamma_{\rm L}$, where $\omega_{\gamma}$ is the photon energy and $\gamma_{\rm L}$ is the Lorentz factor of the colliding nuclei. Therefore the lepton pairs have a very small $p_{\rm T,ll}$ and the two leptons are nearly back-to-back. ATLAS quantifies the deviation from back-to-back in terms of the acoplanarity ($\alpha$) defined as $1 - \frac{|\varphi^{+}-\varphi^{-}|}{\pi}$ where $\varphi^{+}$ and $\varphi^{-}$ are the azimuthal angles of the two muons. Both experiments show a significant broadening of the $p_{\rm T,ee}$ (STAR) or $\alpha$ (ATLAS) distributions of the lepton pairs increasing for more central collisions in HHIC compared to UPCs. Whereas STAR attributed it to the possible deflection of the leptons by a magnetic field trapped in an electrically conducting QGP, ATLAS estimated that the observed broadening is qualitatively consistent with potential electromagnetic scatterings of the leptons with the hot and dense medium. Nevertheless, theoretical models tackling the relationship between $b$ and the transverse momentum of the quasi-real photons were not readily available at the time of those results.

In the past, two main approaches have been used to calculate the photon--photon interactions: the Equivalent Photon Approximation (EPA)\,\cite{Fermi:1925fq,vonWeizsacker:1934nji,Williams:1934ad} and lowest-order QED calculations (LOQED)\,\cite{Hencken:1994my,Alscher:1996mja}. In the EPA framework, the cross section of the two-photon process in heavy-ion collisions is obtained as a folding of the equivalent number of quasi-real photons $n_{\rm 1}(\omega_{\rm \gamma,1})$ and $n_{\rm 2}(\omega_{\rm \gamma,2})$  from the field of the nucleus 1 and 2, respectively, and the elementary photoproduction cross section $\sigma_{\gamma\gamma\to{\rm l^{+} l^{-}}}$. The latter is given by the polarization-averaged cross section of the Breit--Wheeler process. Originally, the $k_{\rm T}$-factorisation method as defined in Refs.\,\cite{Klusek-Gawenda:2018zfz,Klusek-Gawenda:2020eja} was used to calculate the transverse momentum ($k_{\rm T}$) of the quasi-real photons. In such an approach, the shape of the $k_{T}$-photon distribution is assumed to be independent of the collision impact parameter. Measurements of photon--photon produced dileptons by ALICE\,\cite{ALICE:2013wjo}, CMS\,\cite{CMS:2018erd} and ATLAS\,\cite{ATLAS:2020epq} in UPCs are relatively well reproduced by calculations based on the EPA as implemented e.g.\ in STARlight\,\cite{Klein:2016yzr}. Nevertheless, more differential measurements in UPCs show a broadening of the azimuthal back-to-back dilepton correlations or $p_{\rm T,ll}$ distributions, as well as differences in the invariant mass spectra with increasing number of neutrons at forward rapidity in the events\,\cite{STAR:2019wlg,CMS:2020skx,ATLAS:2020epq}. The latter enables the selection of collisions occuring at small $b$ that contain exclusive dileptons in conjunction with the excitation and dissociation of the passing nuclei. On one hand, ATLAS reported that their data can be described by EPA calculations using the $k_{\rm T}$ factorisation approach, as long as an additional, similarly factorized, dissociative contribution is included. In these dissociative processes, one photon is emitted by charged constituents of a nucleon, corresponding to an incoherent component of the photon fluxes. Its contribution was estimated by ATLAS by fitting the measured acoplanarity distributions\,\cite{ATLAS:2020epq}. On the other hand, CMS showed that their results for small $\alpha$ ($\alpha < 0.01-0.02$) can be qualitatively reproduced by LOQED calculations neglecting such dissociative processes but incorporating a $b$ dependence of the shape of the initial photon $k_{\rm T}$\,\cite{CMS:2020skx}. These calculations\,\cite{Brandenburg:2020ozx,Zha:2018tlq} predict a $k_{\rm T}$ hardening of the initial-state photons with a decrease of $b$ as a consequence of the spatial distribution of the EM fields. Attempts to implement $b$ dependences in a generalized EPA approach have been performed in Refs.\,\cite{Vidovic:1992ik,Hencken:2004td,Zha:2018tlq,Brandenburg:2020ozx}. Such calculations show strong impact parameter dependences of the dilepton $p_{\rm T,ll}$ distributions but produce an unphysical increase of the cross section at very low $p_{\rm T,ll}$\,\cite{Zha:2018tlq}, related to neglected interference terms. Recently, an approach using the Wigner formalism suggested in Ref.\,\cite{Li:2019sin} and performed in Refs.\,\cite{Klein:2020jom,Klusek-Gawenda:2020eja,Wang:2021kxm}, was shown to recover the full $b$ dependence of the lowest-order QED calculations.

After including the $b$ dependence of the photon $k_{\rm T}$ distribution in the calculations, the existing results of STAR\,\cite{STAR:2019wlg,STAR:2004bzo,STAR:2018ldd}, ATLAS\,\cite{ATLAS:2020epq,ATLAS:2018pfw,ATLAS:2022vbe}, and  CMS\,\cite{CMS:2020skx} in UPC and HHIC are reasonably well described by LOQED predictions and calculations based on the EPA within the uncertainties of the data. As a consequence, room for any medium-induced or final-state effect in HHIC is significantly reduced, whereas photon--photon interactions turn out to be useful for mapping the EM fields generated by the highly Lorentz-contracted nuclei. Further properties of the $\gamma\gamma\to{\rm e^{+}{\rm e^{-}}}$ process were measured by STAR. In particular, a cos$(4\Delta\varphi)$ angular modulation, where $\Delta\varphi$ is the azimuthal angle in the laboratory frame between the momentum of the \ee pair and one of the electrons, was predicted due to the initial linear photon polarization\,\cite{Li:2019yzy,Li:2019sin}. This feature was confirmed by STAR measurements in UPCs and peripheral Au--Au collisions with hadronic overlap at $\sqrt{s_{\rm NN}}$ $=$ 200~\GeVc\,\cite{STAR:2019wlg} and is closely related to the phenomenon of birefringence\,\cite{Heisenberg:1936nmg}. 

Despite the overall good description of the data by the latest calculations, some points deserve further theoretical and experimental investigation, see Ref.\,\cite{Brandenburg:2021lnj} for an overview. Among them, the effect of higher-order corrections in the QED predictions is unclear\,\cite{Baur:2008hn,Zha:2021jhf}. Due to the large charge carried by the heavy ion, the parameter of the perturbative expansion in such calculations is large. The large tails observed in the measured $p_{\rm T,ee}$ and $\alpha$ distributions could be related to next-leading-order contributions from final state radiation as shown in\,\cite{CMS:2020skx}. With ALICE, the $\gamma\gamma\to{\rm e^{+}}{\rm e^{-}}$ process can be studied in a similar region of phase space as measured by STAR, but in collisions with a much larger Lorentz-boost factor ($\gamma^{\rm LHC}_{\rm L}$ $\approx$ 2700, $\gamma^{\rm RHIC}_{\rm L}$ $\approx$ 100). The maximum electric field reached in heavy-ion collisions is of the order of $Ze\gamma_{\rm L}$/$d^{\rm 2}$\,\cite{Baur:2008hn}, where $d$ the distance from the ion's center, and is consequently about 30 times larger at the LHC compared to RHIC. The fields vary and act over a short timescale of approximately $d$/($\gamma_{\rm L} c$), i.e.\ 10$^{-25}$ (10$^{-23}$) s at the LHC (RHIC). Therefore, measurements of photon--photon production of dielectrons at the LHC would allow the predicted photon kinematic distributions to be experimentally verified for larger expected magnetic fields than at RHIC and could provide further constraints on the mapping of the EM fields produced in heavy-ion collisions, as well as possible medium effects.

In this article, the first measurement of \ee pairs at low \ptee and \mee at the LHC is presented in peripheral (70--90\%) and semi-peripheral (50--70\%) \PbPb collisions at \fivenn. The dielectron production is measured with ALICE at midrapidity (\etarangee{0.8}) and \ptee $<$ 0.1~\GeVc from an invariant mass of 2.7~\GeVmass down to 0.4~\GeVmass. The latter is determined by the minimum \pt required to identify electrons (\pte $>$ 0.2~\GeVc) in the central barrel. The data are compared with the expected dielectron rate from known hadron decays, called the hadronic cocktail, with predictions for thermal radiation from the medium and with recent predictions for coherent photoproduction of dielectrons as a function of \mee. The \ptee and  \pteesquare distributions are extracted in three different \mee ranges in peripheral \PbPb collisions and the extracted value of $\sqrt{\langle \ptee^{2} \rangle}$ is compared with predictions and to measurements at lower \snn. 

The article is organized as follows. Section~\ref{Detector} contains a brief description of the ALICE apparatus and the data sample used, whereas Section~\ref{analysis} illustrates the analysis steps. In Section~\ref{results}, the results on dielectron production yields at low \ptee within the ALICE acceptance are presented and compared with theoretical calculations and previous measurements at lower \snn. Section~\ref{summary} gives a summary and outlook.

\section{Detector and data samples}
\label{Detector}

A detailed description of the ALICE apparatus and its performance can be found in Refs.\,\cite{ALICE:2008ngc,ALICE:2014sbx}. The main detectors used to track and identify electrons\footnote{Note that the term ‘electron’ is used for both electrons and positrons throughout this paper.} at midrapidity (\etarangee{0.8}) are the Inner Tracking System (ITS)\,\cite{ALICE:2010tia}, the Time Projection Chamber (TPC)\,\cite{Alme:2010ke}, and the Time-Of-Flight (TOF) detector\,\cite{Akindinov:2013tea}. The ITS consists of six cylindrical layers of silicon detectors, which provide tracking of the charged particles and, together with the TPC, the reconstruction of the primary collision vertex. The innermost layer is installed at a radius of 3.9\,cm from the beam axis and is used to reject electrons from photon conversions in the detector material. The TPC detector allows tracks to be reconstructed and charged particles to be identified (PID) via the measurement of the specific energy loss d$E$/d$x$ while the TOF detector contributes to the PID via the measurement of the flight time of the particles. These detectors are placed inside a uniform magnetic field of 0.5\,T parallel to the beam direction, provided by a solenoid magnet.

The data samples used in this analysis were collected by ALICE in 2015 and 2018 during \PbPb runs at \fivenn. Minimum-bias collisions were triggered by requiring the coincidence of signals in the two scintillator arrays of the V0 detectors\,\cite{Abbas:2013taa}, covering the pseudorapidity ranges $2.8 \leq \eta < 5.1$ and $-3.7 \leq \eta < -1.7$. The time information from the V0 detectors and the neutron Zero Degree Calorimeters (ZDC)\,\cite{Arnaldi:1999zz}, as well as the correlation between the number of hits in the ITS and in the TPC are used offline to reduce the background from beam--gas interactions and  pile-up collisions to a negligible level. Only events with a primary vertex reconstructed close to the center of ALICE along the beam direction ($|z| < 10$\,cm) are considered in the analysis to assure a uniform detector acceptance. The event sample was divided into centrality classes\,\cite{ALICE-PUBLIC-2018-011} expressed in percentages of the total hadronic cross section using the amplitudes of the signal in the V0 detector. The number of events in each centrality class considered in this analysis, i.e.\ 50--70\% and 70--90\%, is about 34 million after the event selection criteria.

\section{Data analysis}
\label{analysis}

\subsection{Electron candidate selection}

Electron candidates are selected from charged-particle tracks reconstructed in the ITS and TPC in the kinematic range \etarangee{0.8} and \pte $>$ 0.2~\GeVc. The track fits are required to include at least 80 out of a maximum of 159 reconstructed space points in the TPC and a hit in at least 4 of the 6 ITS detector layers. The $\chi^{\rm 2}$ per space point measured in the TPC (ITS) must be less than 2.5 (5). In order to reduce the contribution of secondary tracks arising from weak decays and interactions with the detector material, only tracks with a distance-of-closest approach to the reconstructed primary vertex smaller than 1\,cm in the plane transverse to the colliding beams and 0.5\,cm in the longitudinal direction are used in the analysis. In addition, a hit in the first ITS layer is required to reject electrons originating from real-photon conversions in the detector material of the subsequent ITS layers. Since the electrons originating from the same photon conversion share the same cluster in the ITS layer where they are produced, they can be further suppressed by requiring that a maximum of one ITS cluster attached to the reconstructed track is shared with any other track candidate and is not placed in the first ITS layer.

The electron identification is based on the complementary information provided by the TPC and TOF. The detector PID signal, $n$($\sigma^{\rm DET}_{\rm i}$), is expressed in terms of the deviation between the measured and expected value of the specific ionisation energy loss in the TPC or time-of-flight in the TOF for a given particle hypothesis $i$ and momentum, normalised to the respective detector resolution. In the TPC, electrons are selected in the range $|n\rm{(\sigma^{TPC}_{e})}| \leq 3$, whereas kaons, protons and pions are rejected with $|n\rm{(\sigma^{TPC}_{K})}| \geq 3$, $|n\rm{(\sigma^{TPC}_{p})}| \geq 3$ and $n\rm{(\sigma^{TPC}_{\pi})} \geq 3.5$, respectively. Electrons with an energy loss in the TPC in the range where the charged kaon and proton bands cross the one of electrons are recovered using the TOF information: tracks which fulfill only the TPC electron selection and pion rejection but have an associated TOF signal with $|n\rm{(\sigma^{TOF}_{e})}| \leq 3$ are accepted. This PID strategy was used successfully in previous ALICE dielectron analyses in \pp and \pPb collisions\,\cite{ALICE:2020mfy,ALICE:2020umb,ALICE:2018gev}. Averaged over $p_{\rm T}$, the hadron contamination in the single-electron candidate sample is less than 5\% for an electron efficiency of about 80\%. The largest hadron contamination, up to about 18\% in the 50--70\% centrality class, is observed where kaons ($p_{\rm T} \approx 0.5$~\GeVc), protons ($p_{\rm T} \approx 1$~\GeVc), or charged pions ($p_{\rm T} > 6$~\GeVc) have a similar d$E$/d$x$ as electrons in the TPC. Pairs containing a misidentified hadron are further removed during the subtraction of the combinatorial background, thus that the final hadron contamination in the dielectron signal is expected to be negligible.

\subsection{Signal extraction}

Electron pairs originating from the same source cannot be identified unambiguously. Therefore, a statistical approach is used to extract the yield of signal pairs ($S$), in which all electrons and positrons in an event are combined to create an opposite charge-sign spectrum ($OS$). The combinatorial background ($B$) is estimated from same-event pairs with the same charge sign ($SS$). In comparison to a mixed-event approach~\cite{PHENIX:2009gyd}, the same charge-sign approximation of the combinatorial background has the advantage to be self-normalized and to contain all residual correlations arising from charge-symmetric processes, such as from conversions of correlated decay photons originating from the same and from decays of different hadrons inside the same jets or in back-to-back jets. A different acceptance for opposite charge-sign and same charge-sign pairs is observed arising from detector geometrical effects, i.e.\ non-uniformity of the detector performances in azimuthal angle $\varphi$. The correction factor $R_{\rm acc}$, needed to account for this effect, is calculated with an event-mixing technique detailed in Ref.\,\cite{ALICE:2018ael}. Events with similar global properties are grouped together according to the $z$-position of the reconstructed primary vertex, the centrality of the collision, and the event-plane angle estimated with the V0 detector. The factor $R_{\rm acc}$ is found to be consistent with unity above \mee of 1~\GeVmass. The signal is then extracted as $S = OS - R_{\rm acc} \times SS$.  

\begin{figure}[tbh]
    \begin{center}
    \includegraphics[width = 0.495\textwidth]{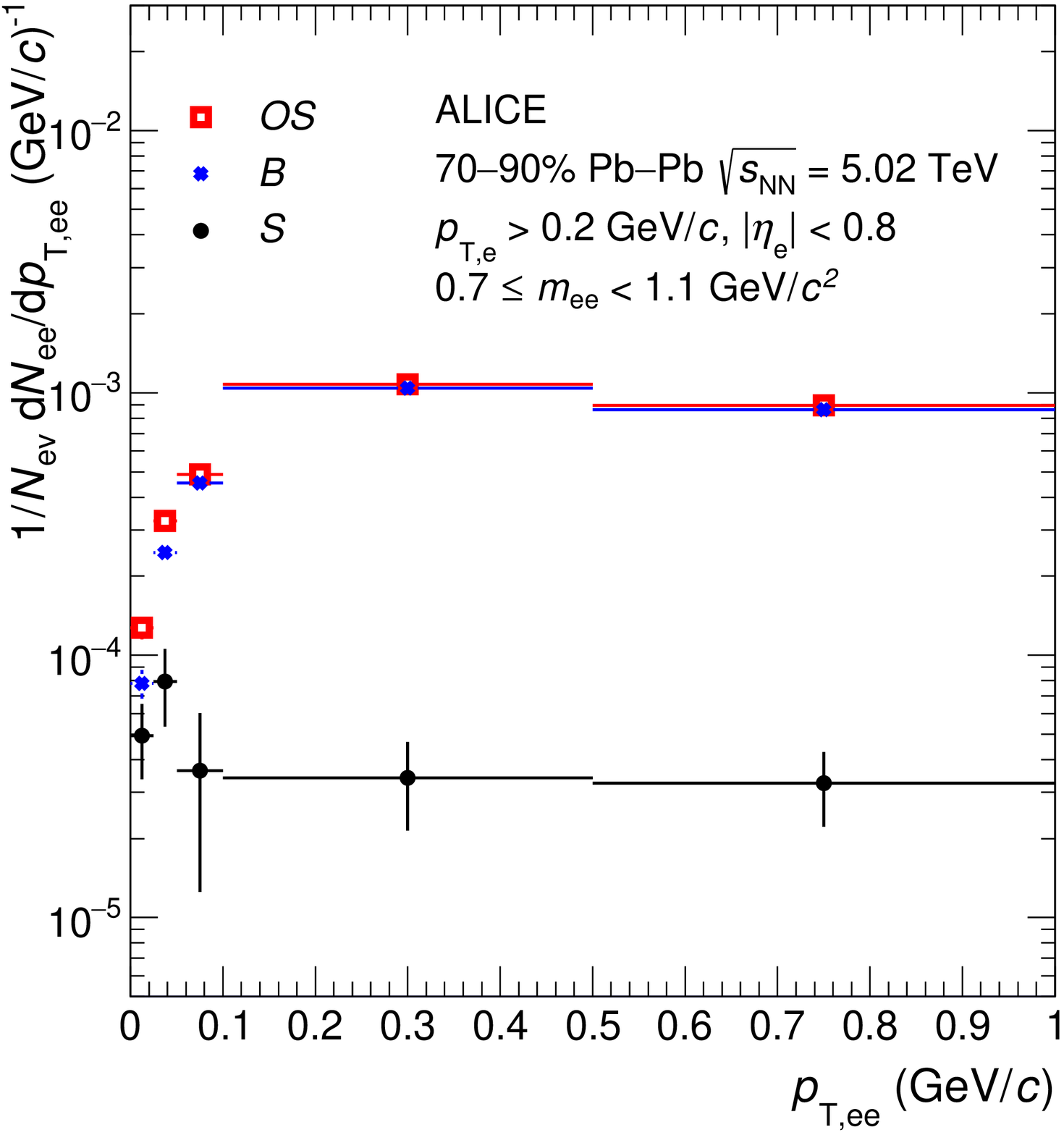}
    \includegraphics[width = 0.495\textwidth]{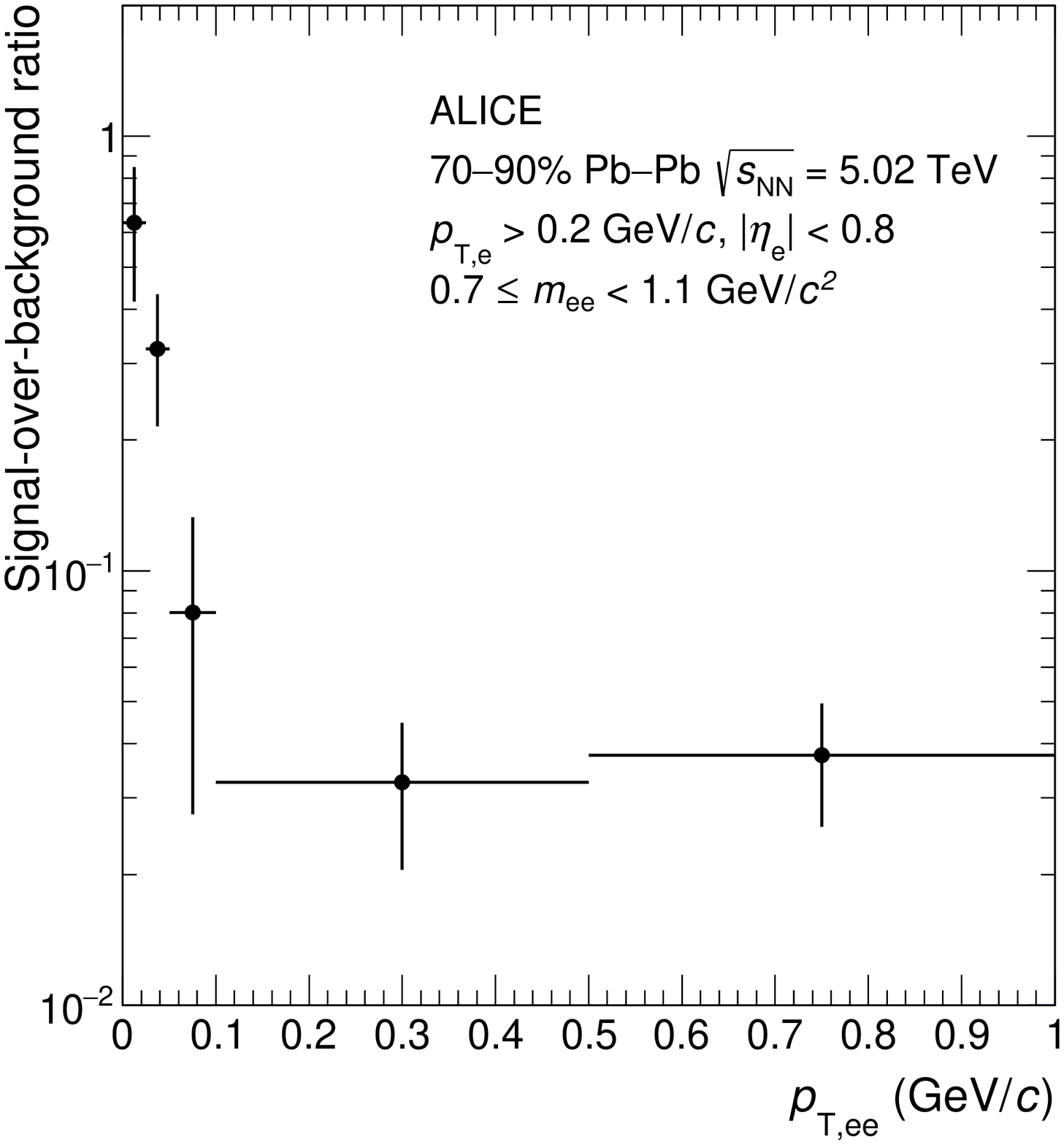}      
    \end{center}
    \caption{Left panel: raw $p_{\rm T,ee}$-differential yield ($S$) in peripheral (70--90\%) Pb--Pb collisions at \fivenn for $0.7 \leq \mee < 1.1$~\GeVmass overlaid with the opposite charge-sign distribution ($OS$) and the same charge-sign spectrum multiplied by the acceptance correction factor $R_{\rm acc}$ ($B$). Right panel: signal over background as a function of \ptee in peripheral (70--90\%) Pb--Pb collisions at \fivenn for $0.7 \leq \mee < 1.1$~\GeVmass.}
    \label{fig:signalextraction}
\end{figure}

The opposite charge-sign spectrum, the combinatorial background, and the extracted raw dielectron signal are shown in the left panel of Fig.~\ref{fig:signalextraction} as a function of the pair transverse momentum \ptee for \mbox{$0.7 \leq\mee < 1.1$~\GeVmass} in 70--90\% peripheral Pb--Pb collisions. The corresponding signal-over-background ratio ($S$/$B$) is presented in the right panel of Fig.~\ref{fig:signalextraction}. Towards very low \ptee (\ptee $\leq$ 0.1~\GeVc), the $S$/$B$ ratio increases for both centrality classes. However, the $S$/$B$ ratio is about one order of magnitude lower in the 50--70\% centrality class in this \ptee region.   

\subsection{Efficiency correction}

The raw signal is corrected for the finite dielectron reconstruction efficiency. To this end, different Monte Carlo (MC) simulations are used, where a realistic detector response is modelled using GEANT3\,\cite{Brun:1119728}. For very low \ptee ($\ptee < 0.2$~\GeVc), photoproduced \ee pairs are simulated with the event generator STARlight\,\cite{Klein:2016yzr} and embedded into hadronic collisions computed with HIJING\,\cite{Wang:1991hta}. At larger \ptee, additional samples of dielectron sources injected into HIJING simulated events are utilized. These include light-flavour hadrons ($\pi^{\rm 0}$, $\eta$, $\eta\prime$, $\rho^{\rm 0}$, $\omega$ and $\phi$) and $J$/$\psi$ mesons, forced to decay into dielectrons with the phenomenological EXODUS generator\,\cite{PHENIX:2009gyd} and PHOTOS\,\cite{Golonka:2005pn}, respectively, and produced in equal amounts with uniform \pt distributions. In each centrality class (50--70\% or 70--90\%), these input \pt distributions are corrected with \pt-dependent weights defined as the ratio of the hadron \pt spectra in the MC simulations and the expected hadron \pt  distributions according to the hadronic cocktail explained in Section~\ref{hadroniccocktail}. The weights are passed to the decay electrons to produce a realistic mix of \ee pairs from the various sources considered. In addition, an enriched sample of heavy-flavour hadron sources with enforced semileptonic decay channels generated with the Perugia 2011 tune of PYTHIA 6.4\,\cite{Sjostrand:2006za,Skands:2010ak} is used. The final efficiency as a function of \mee and \ptee is the average of the efficiencies of the different dielectron sources, weighted by their expected contribution, for \ptee $\geq$ 0.2~\GeVc. At lower \ptee only the STARlight calculations are taken as input. Other sources show dielectron efficiencies in agreement within statistical uncertainties with the one extracted for \ee pairs produced via photon--photon interactions.

\subsection{Systematic uncertainties of measured dielectron spectra}

The systematic uncertainties on the measured \ptee- and \mee-differential dielectron yields in peripheral (70-90\%) and semi-peripheral (50-70\%) collisions originate from tracking, electron identification and purity, and background subtraction. They are evaluated as described in Ref.\,\cite{ALICE:2020mfy} and summarised in Table~\ref{systable} for $\ptee <  0.1$~\GeVc.

\begin{table}[ht!]
\begin{center}
\centering
\caption{Summary of the total systematic uncertainties of the measured dielectron yields for $p_{\rm T,ee} <$ 0.1~GeV/$c$ in semi-peripheral (50--70\%) and peripheral (70--90\%) \PbPb collisions at \fivenn. The values presented as a range correspond to the smallest and largest observed systematic uncertainties.}
\begin{tabular}{lllllllllll}
 \hline 
  Centrality    & Hit in the  &  TPC--TOF & ITS--TPC &   Shared  & Tracking & Anchor &    Total  \\
             class        & first ITS layer  & matching & matching &   ITS cluster & and PID &  point &    \\
\hline
50--70\%  & 2\%   &  0--4\%  & 5.4--7.4\%    &   4\%    & 16\% & 0\% &  18\% \\\hline
70--90\%  & 2\%   &  0--4\%  &  5.4--7.4\%    &   4\%    & 6\% & 5\% &    10--12\% \\\hline                                                                
\hline 
\end{tabular}
\label{systable}
\end{center}
\end{table}

The systematic uncertainties related to the requirement of a hit in the innermost ITS layer, the matching of the TPC track and the signal measured in the TOF, and the matching of the track segments reconstructed in the ITS and the TPC are first estimated at the single-track level. To this end, the efficiencies of these selection criteria are compared in data and in MC as a function of \pt for a pure sample of charged pions or electrons (TPC--TOF matching). The latter is obtained by selecting electrons from photon conversions in the detector material using topological requirements. A MC method is then used to calculate the corresponding uncertainties for dielectrons, by generating particles in the full \mee and \ptee phase space and forcing them to decay to \ee pairs. The uncertainty for each \ee pair is given by the sum of the uncertainties of the decay electrons, after applying the fiducial selection (\etarangee{0.8} and $\pte \geq 0.2$~\GeVc). The final systematic uncertainty is obtained after averaging  for a given \mee and \ptee over all generated particles. The TPC--TOF matching efficiency is relevant only in the regions where the kaon and proton bands cross the band of electrons in the TPC. The corresponding uncertainty varies between 0 and 4\% for the \ee pairs and is the largest for the invariant mass bin $1.1 \leq \mee < 2.7$~\GeVmass at low \ptee ($\ptee < 0.1$~\GeVc). The ITS--TPC matching efficiency is one of the dominant sources of systematic uncertainties together with the particle identification and leads to uncertainties between 5.4\% and 7.4\% increasing with \mee. The systematic uncertainty originating from the requirement of a hit in the first ITS layer is of the order of 2\%.

The systematic uncertainty from the requirement on the number of ITS shared clusters is estimated by varying the number of allowed shared ITS clusters for the selected electron candidates and repeating the analysis steps. Releasing completely this selection criterion increases significantly the amount of electrons from conversions in the detector material and leads to a smaller S/B by a factor of about 0.6. Therefore the extracted systematic uncertainty contains not only systematic effects from the signal efficiency, but also from the background estimation. It is calculated from the maximum deviations of the efficiency-corrected spectra variations, considered as statistically significant according to the Barlow criterion\,\cite{Barlow:1993dm} and found to be of the order of 4\%.

In a similar way, the systematic uncertainty arising from the tracking and electron identification and purity is evaluated by varying the remaining electron selection criteria simultaneously, e.g. the requirement on the minimum number of reconstructed space points in the TPC or $|n\rm{(\sigma^{TPC}_{e})}|$, to take into account possible correlations between them. In particular modifying the requirements on the TPC and TOF signals, i.e.\ $|n\rm{(\sigma^{TPC}_{\rm e})}|$,  $|n\rm{(\sigma^{TPC}_{\rm \pi})}|$,  $|n\rm{(\sigma^{TPC}_{\rm K})}|$,  $|n\rm{(\sigma^{TPC}_{\rm p})}|$ and  $|n\rm{(\sigma^{TOF}_{\rm e})}|$, enables to probe possible biases due to differences in the detector responses in data and MC and remaining hadron contamination in the electron sample. The systematic uncertainty is computed as the root-mean-square of the variation of the final data points and is found to be of the order of 16\% (6\%) in semi-peripheral (peripheral) Pb--Pb collisions for $\ptee < 0.1$~\GeVc. The main source of systematic uncertainty in the 50--70\% centrality class comes from the kaon and proton rejection in the TPC and the non-perfect description of the measured particle energy loss in the TPC in the simulations, which depends on the centrality of the collisions.

The systematic uncertainty originating from the correction factor $R_{\rm acc}$, estimated by varying the event mixing pools used to calculate it, was found to be negligible at low \ptee.

Finally, systematic uncertainties arise from the centrality class definition. The absolute scale of the centrality is defined by the range of 0--90\% centrality in which a Glauber-based multiplicity model is fitted to the V0M distribution\,\cite{ALICE-PUBLIC-2018-011}. The lower centrality limit of 90\% of this range with its corresponding V0M signal is denoted the anchor point (AP). The AP was shifted by $\pm 1$\%, leading to a systematic uncertainty of 5\% for the 70--90\% centrality class and negligible for the 50--70\% centrality class.

\subsection{Expected yield from known hadronic sources}
\label{hadroniccocktail}

The expected dielectron yield from the decays of known hadrons produced in the hadronic \PbPb collisions, called the hadronic cocktail, is calculated with a fast simulation of the ALICE central barrel, including the angular and momentum resolution of the detector and bremsstrahlung effects\,\cite{ALICE-PUBLIC-2017-011}. 

The Dalitz and dielectron decays of light neutral mesons are simulated following the approach described in Ref.\,\cite{ALICE:2018fvj}. The \pt-differential production cross sections of $\eta$ and $\omega$ are estimated based on the ratio of their \pt spectra to the one of $\pi^{0}$ or $\pi^{\pm}$, measured in different collision systems and at different center-of-mass energies, whereas $\eta^{\prime}$, $\rho$, and $\phi$ are generated assuming $m_{\rm T}$-scaling over the full \pt range or only at low \pt\,\cite{WA80:1995whm,Khandai:2011cf,Altenkamper:2017qot}. The \pt spectra of $\pi^{\pm}$, measured down to a \pt of 0.1~\GeVc as a function of the collision centrality in \PbPb collisions at \fivenn\,\cite{ALICE:2019hno}, are parametrized and extrapolated to \pt $=$ 0 using a two-component function\,\cite{Bylinkin:2012bz,Bylinkin:2015xya}. The difference between $\pi^{0}$ and $\pi^{\pm}$ due to isospin-violating decays is taken into account using an effective model that describes measured hadron spectra ($\pi^{\pm}$, K$^{\pm}$, and p\,\cite{ALICE:2019hno}) at low \pt and includes strong and electromagnetic decays\,\cite{Reygers:2019aul}, as described in Ref.\,\cite{ALICE:2020umb}. This leads to \pt-dependent scaling factors applied to the $\pi^{\pm}$ parametrizations of about 1.3 for \pt $\to$ 0 and consistent with unity within 2\% for \pt $>$ 1~\GeVc. The \pt spectrum of $\eta$ is computed as the average of the spectra obtained using the parametrizations retrieved from the $\eta$/$\pi^{0}$ ratio as a function of \pt in \pp collisions\,\cite{ALICE:2020umb} and from the K$^{\pm}$/$\pi^{\pm}$ ratio as a function of \pt measured down to \pt $= 0.3$~\GeVc in \PbPb collisions\,\cite{ALICE:2019hno}. In all considered centrality classes (50--70\% and 70--90\%), the ratio of the resulting \pt distribution of $\eta$ to the $\pi^{0}$ parametrization at very low \pt \mbox{($\pt \leq 0.1$~\GeVc)} was found to be in agreement within uncertainties with the $\eta$/$\pi^{0}$ ratio in \pp collisions. The latter is constrained at low \pt by the data from CERES/TAPS\,\cite{Agakichiev:1998ign} and has a conservative \pt-dependent uncertainty of up to 40\%, which is taken into account in the final uncertainty of the hadronic cocktail. At \mee around 0.782~\GeVmass, the dominant contribution to the hadronic cocktail is given by the $\omega$ meson. A parametrization of the $\omega$/$\pi^{\rm 0}$ ratio as a function of \pt measured by ALICE in \pp collisions at \seven\,\cite{ALICE:2020ylo} is performed and extended to \pt $=$ 0 using data from PHENIX in \pp collisions at $\sqrt{s} = 200$~GeV\,\cite{PHENIX:2011sbe}. It is used for all centrality classes. Finally, the measured \pt spectra of $\phi$ mesons in semi-central and peripheral \PbPb collisions at \fivenn\,\cite{ALICE:2019xyr} are fitted and extrapolated down to low \pt (\pt $\leq$ 0.4~\GeVc) using $m_{\rm T}$ scaling to obtain the $\phi$ input parametrizations.

The contribution from correlated semileptonic decays of open charm and beauty hadrons is computed with the next-to-leading order event generator POWHEG~\cite{powheg1,powheg2,powheg3,powheg4} with PYTHIA 6\,\cite{Sjostrand:2006za} to evolve the parton shower. The expected yield is normalized to the  cross sections $\rm{d}\sigma_{\rm c\bar{c}}/\rm{d}y|_{y=0}$ and $\rm{d}\sigma_{\rm b\bar{b}}/\rm{d}y|_{y=0}$ extracted with the same MC generator from the \ee spectra measured in \pp collisions at $\sqrt{s} = 5.02$~TeV\,\cite{ALICE:2020mfy} and scaled with the nuclear overlap function. The resulting contribution from correlated open heavy-flavour hadron decays dominates the hadronic cocktail yield for $\ptee < 0.1$~\GeVc up to \mee of 2.7~\GeVmass, except in the mass regions around 0.4, 0.78 and 1.~\GeVmass, where the $\eta$, $\omega$ and $\phi$ are the main sources of \ee pairs, respectively. The uncertainties related to the branching ratio of the semileptonic decays of the open heavy-flavour hadrons and the fragmentation functions of charm and beauty quarks are omitted under the assumption that these do not change from \pp to peripheral and semi-peripheral \PbPb collisions. 

The systematic uncertainties of the hadronic cocktail are computed by adding in quadrature the uncertainties originating from the following sources: the $\pi^{\pm}$ and $\phi$ parametrizations as a function of \pt, the $\pi^{\rm 0}$/$\pi^{\pm}$ correction factor, the $\eta$/$\pi^{\rm 0}$ and $\omega$/$\pi^{0}$ ratios, the $m_{\rm T}$-scaling parameters used for $\eta^{\prime}$, $\rho$ and $\phi$, the branching ratios of the different light-flavour hadron decay channels, the heavy-flavour cross sections and the nuclear overlap function. The final systematic uncertainty of the hadronic cocktail at very low \pt ($\ptee < 0.1$~\GeVc) is between 14\% in the intermediate mass range ($1.1 \leq \mee < 2.7$~\GeVmass) and about 30\% in the mass regions dominated by $\eta$ and $\omega$ decays.

\section{Results}
\label{results}

\subsection{Invariant mass spectra}

\begin{figure}[tbh]
    \begin{center}
    \includegraphics[width = 0.6\textwidth]{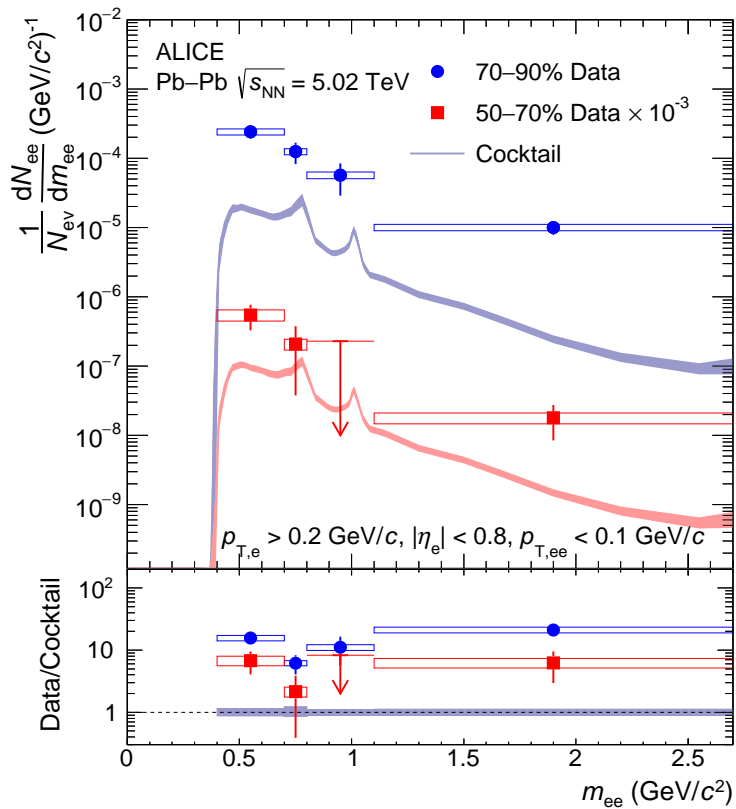}
    \end{center}
    \caption{Dielectron $m_{\rm ee}$-differential yields in semi-peripheral (50--70\%) and peripheral (70--90\%) Pb--Pb collisions at \fivenn, compared with the expected e$^{+}$e$^{-}$ contributions from known hadronic decays. The error bars and boxes represent the statistical and systematic uncertainties of the data, respectively, whereas the bands show the uncertainties of the hadronic cocktail. Arrows indicate upper limits at 90\% confidence level.}
    \label{fig:invmasswithcocktail}
\end{figure}

The efficiency-corrected \ee invariant mass spectra at low \ptee ($\ptee < 0.1$~\GeVc) are shown in Fig.~\ref{fig:invmasswithcocktail} in peripheral (70--90\%) and semi-peripheral (50--70\%) \PbPb collisions at \fivenn within the ALICE acceptance (\etarangee{0.8} and $\pte > 0.2$~\GeVc). In this figure and the following ones, the upper limit at 90\% C.L. using the Feldman and Cousins
methodology\,\cite{Feldman:1997qc} is reported for the results which are found to be statistically consistent with zero within one standard deviation. The data are compared with cocktails of expected e$^{+}$e$^{-}$ hadronic sources. The corresponding enhancement factors, expressed as ratios of data over hadronic cocktail, are illustrated in the bottom panel of Fig.~\ref{fig:invmasswithcocktail}. The total uncertainty of the cocktail is represented by a band. An excess of dielectrons compared to the hadronic expectation is observed in both centrality classes, with a larger significance in peripheral \PbPb collisions.

\begin{figure}[tbh]
    \begin{center}
     \includegraphics[width = 0.495\textwidth]{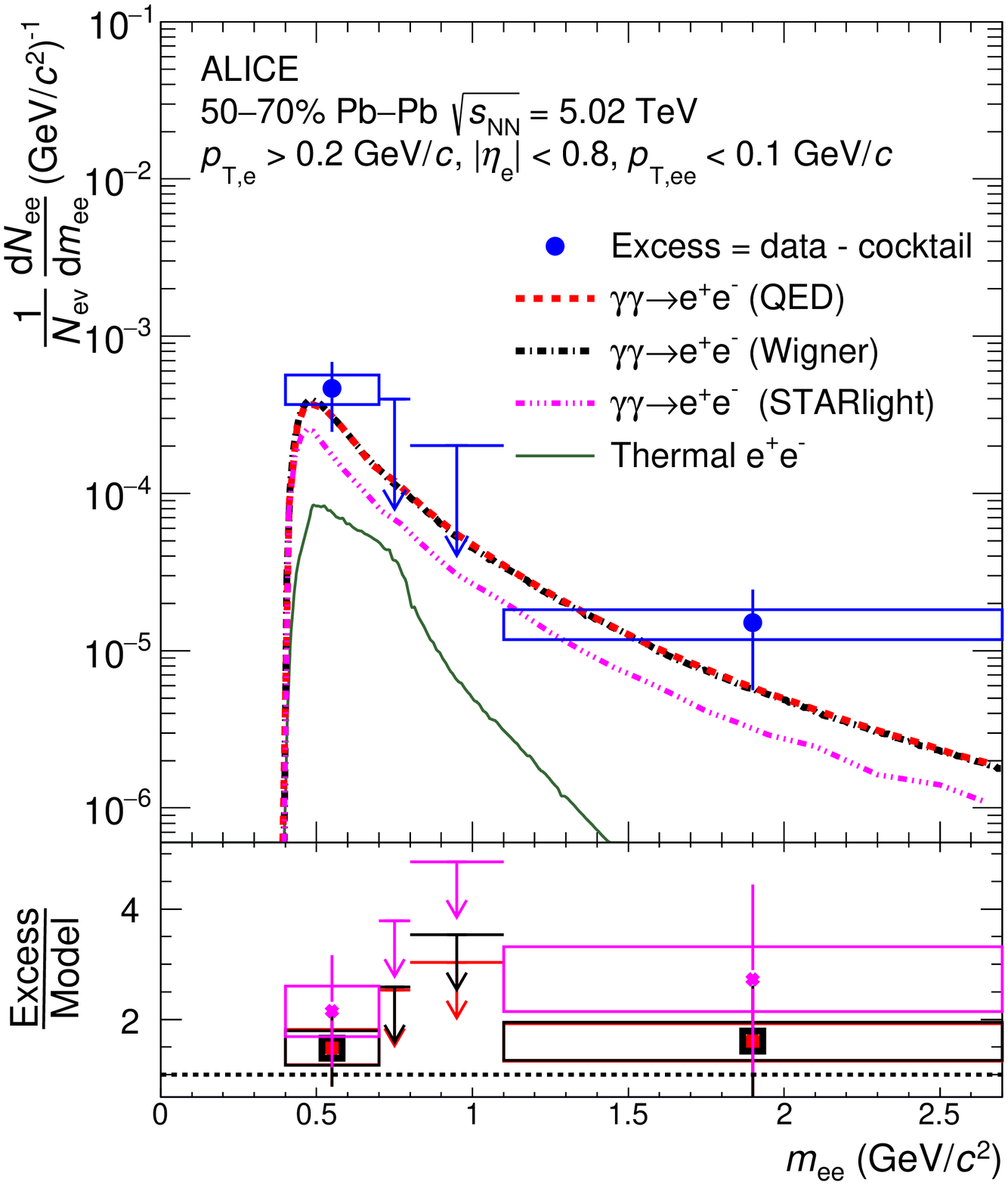}
    \includegraphics[width = 0.495\textwidth]{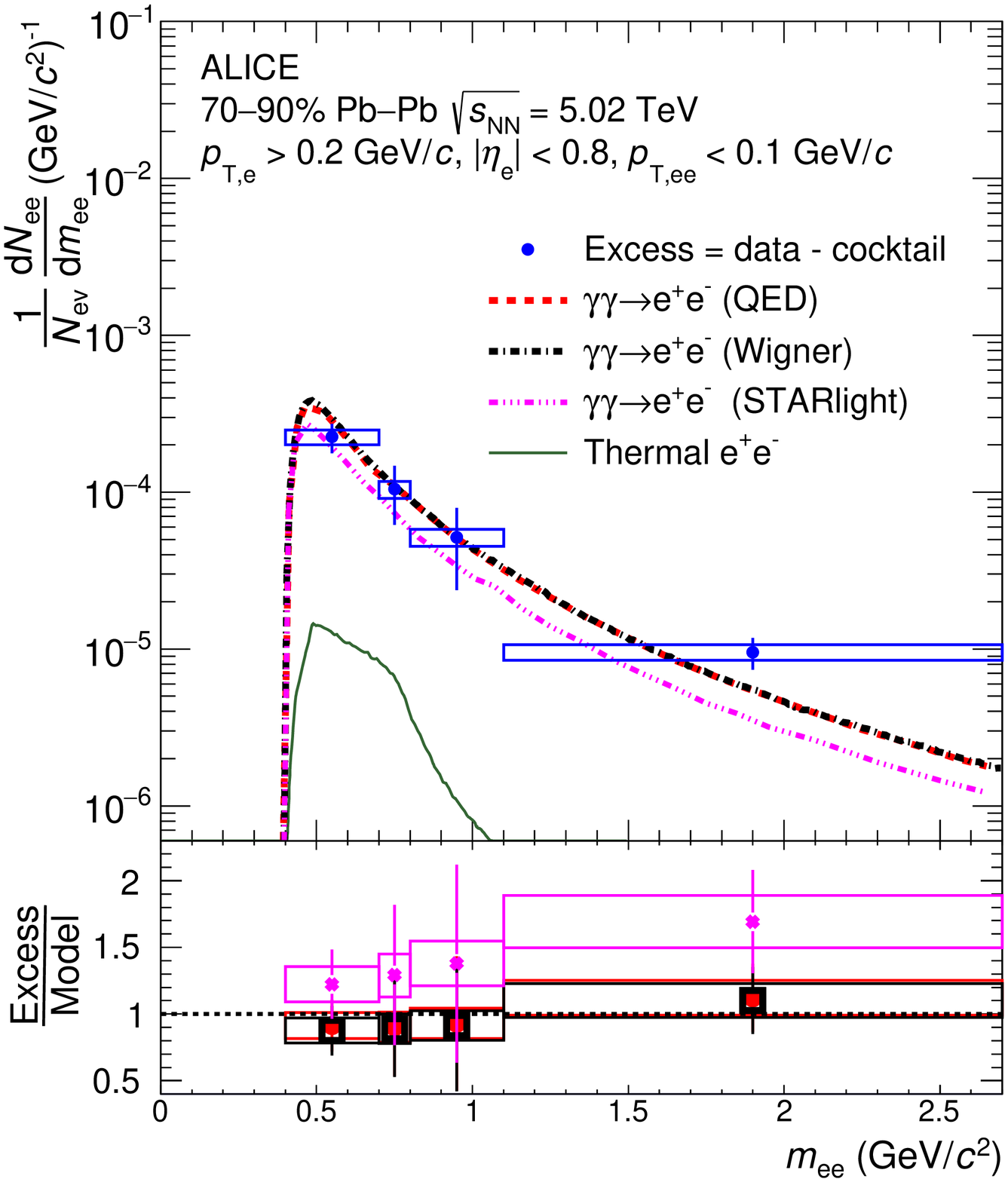}    
    \end{center}
    \caption{Excess dielectron $m_{\rm ee}$-differential yields after subtraction of the cocktail of known hadronic decay contributions in semi-peripheral (left) and peripheral (right) Pb--Pb collisions at \fivenn, compared with calculations for coherent two-photon production of \ee pairs folded with the detector resolution\,\cite{Zha:2018tlq,Brandenburg:2021lnj,Klusek-Gawenda:2020eja,Klein:2016yzr,Klein:2018cjh}. For details see the text. The error bars and boxes represent the statistical and systematic uncertainties of the data, respectively. Arrows indicate upper limits at 90\% confidence level.}
    \label{fig:invmasswiththeory}
\end{figure}

The hadronic cocktail contribution is subtracted from the inclusive \ee pairs to obtain the invariant mass distributions for excess \ee pairs with $\ptee < 0.1$~\GeVc presented in the left and right panels of Fig.~\ref{fig:invmasswiththeory} for the 50--70\% and 70--90\% centrality classes, respectively. The yield of excess \ee pairs does not show a significant centrality dependence. The expected contributions from thermal dielectrons from the partonic and hadronic phases are also shown in the figure. They are estimated with an expanding thermal fireball model including an in-medium broadened $\rho$ spectral function\,\cite{Rapp:1999us,vanHees:2007th,Rapp:2013nxa}. Predictions from the same model describe well the SPS\,\cite{NA60:2006ymb,CERES:2006wcq} and RHIC\,\cite{STAR:2013pwb,PHENIX:2015vek} data. At $\ptee < 0.1$~\GeVc, thermal radiation from the medium is expected to be at least one order of magnitude smaller than the measured \ee excess in peripheral \PbPb collisions and have a different \ptee shape and centrality dependence\,\cite{Klusek-Gawenda:2018zfz}. The excess yield in the \ee invariant mass spectra are further compared with different calculations for photon--photon production of dielectrons. A QED calculation at leading-order was performed by the authors of Refs.\,\cite{Zha:2018tlq,Brandenburg:2021lnj}. The lowest-order two-photon interaction is a second-order process with two contributing Feynman diagrams, as shown in Fig.~2 of Ref.\,\cite{Hencken:1994my}. Higher-order contributions are ignored, although the parameter of the perturbative expansion, the coupling Z$\alpha$ with $\alpha$ the fine structure constant, is close to unity, i.e.\ 0.6, for lead ions. The straight-line approximation for the incoming projectile and target nuclei is applied, as for the other calculations. The predictions from the authors of Ref.\,\cite{Klusek-Gawenda:2020eja} employ the Wigner formalism. The quasi-real photon fluxes originating from strong EM fields produced by the highly Lorentz-contracted heavy ions passing each other can be written in terms of Wigner functions in momentum and impact-parameter space. The cross section for the $\gamma\gamma\to{\rm e^{+}}{\rm e^{-}}$ process is then expressed as a convolution over impact parameters and transverse momenta. Realistic charge form factors of the Pb nuclei, i.e Fourier transforms of the charge density, are taken from Ref.\,\cite{Klusek-Gawenda:2010vqb}. About 50\% of the \ee pairs are produced inside the nuclei for the centrality class 70--90\%. The model implemented in the STARlight MC generator uses the equivalent photon approximation approach\,\cite{Klein:2016yzr,Klein:2018cjh}. The main difference between STARlight and the two aforementioned calculations is related to the treatment of the $b$ dependence in the computations. STARlight utilizes the $k_{\rm T}$-factorisation method, where the one-photon distribution is integrated over all transverse distances to obtain the shape of the $k_{\rm T}$ distribution. For all models, the \mee and \ptee detector resolution, not corrected in the data, are taken into account by folding the momentum and opening angle resolution, including bremsstrahlung effects, in the calculations. As a result, the predicted \mee distributions are slightly softer than the ones computed with perfect detector resolution. The magnitude of the effect is nevertheless below the sensitivity of the data. All models can reproduce the measured $m_{\rm ee}$ excess spectra within their uncertainties. The ratios of the measured excess yields to the different calculations, shown in the bottom panels of Fig.~\ref{fig:invmasswiththeory}, are consistent with unity within the statistical and systematic uncertainties of the data in both centrality classes. However, the STARlight predictions appear to be further away from the data than the other calculations. The contributions from decays of vector mesons produced in photo--nuclear collisions are expected to be very small for $\rho$, $\omega$ and $\phi$\,\cite{STAR:2018ldd,STAR:2019wlg} and below 5\%  based on ALICE results for photoproduced $J/\psi$ at forward rapidity in \PbPb collisions at \fivenn\,\cite{aliceforwardphotojpsi} extrapolated to midrapidity using the IIM model scenario 2 in Ref.\,\cite{GayDucati:2018who}.

\subsection{Transverse momentum spectra}

\begin{figure}[tbh]
    \begin{center}
    \includegraphics[width = 0.329\textwidth]{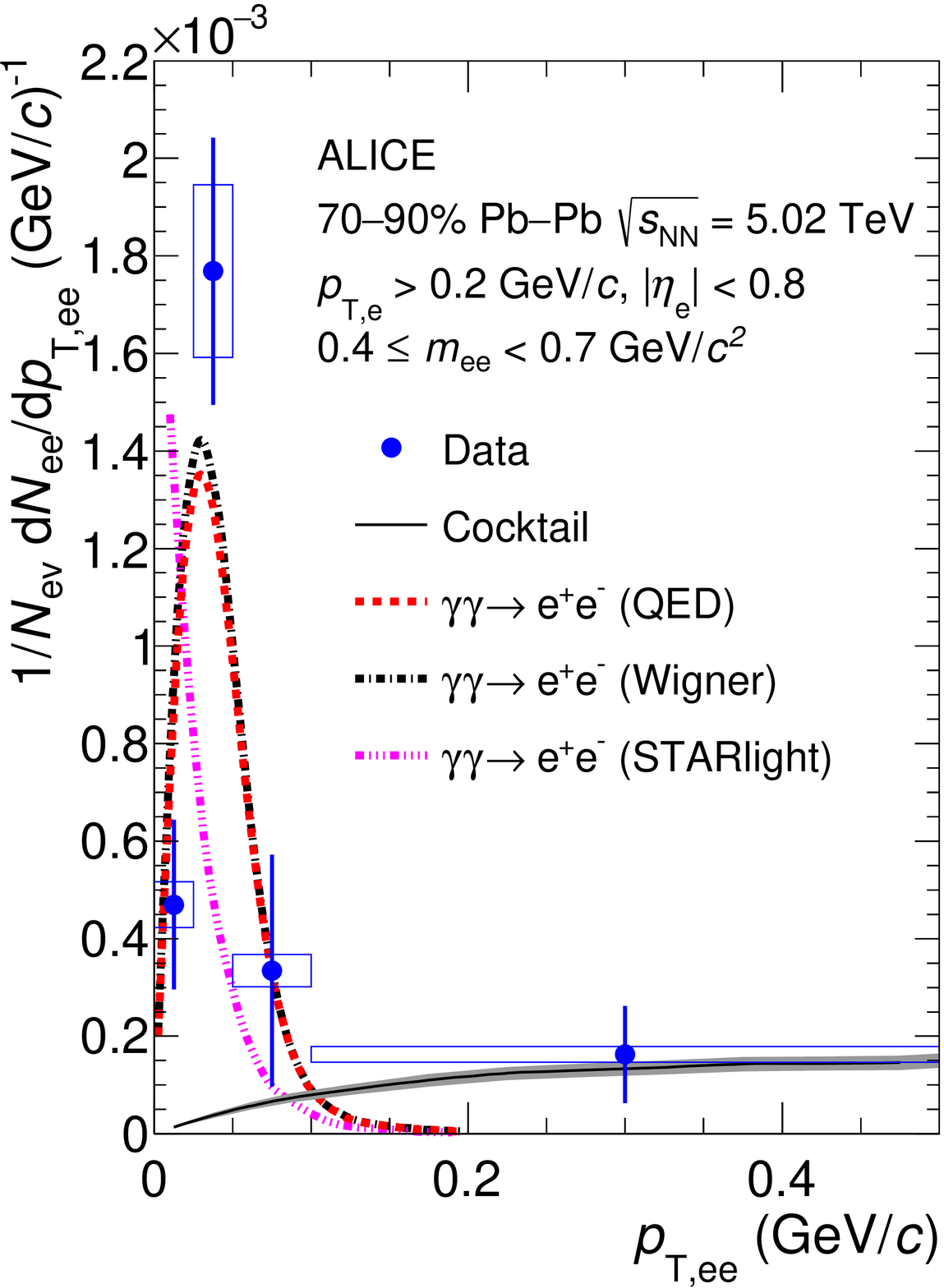}
    \includegraphics[width = 0.329\textwidth]{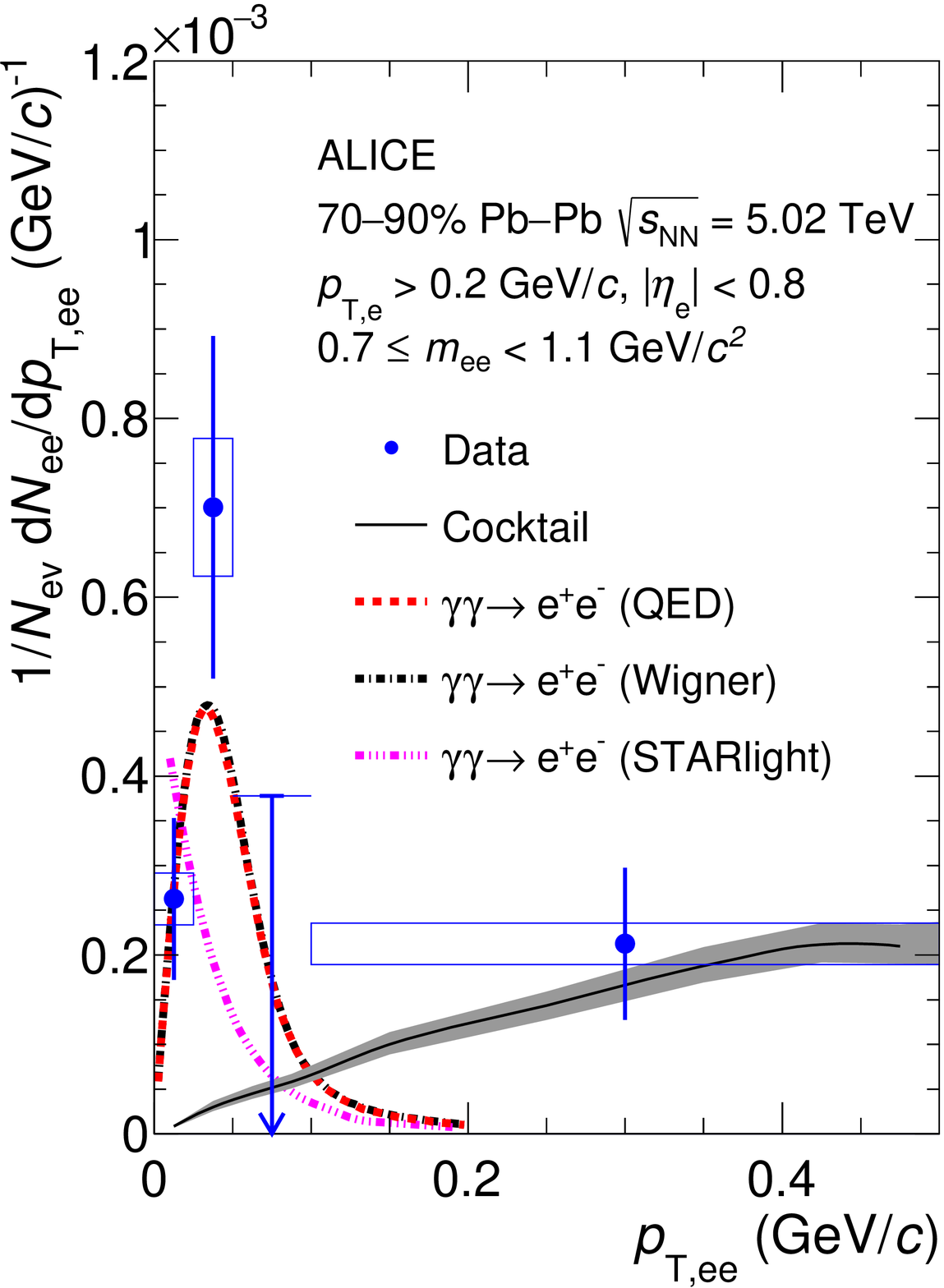}
    \includegraphics[width = 0.329\textwidth]{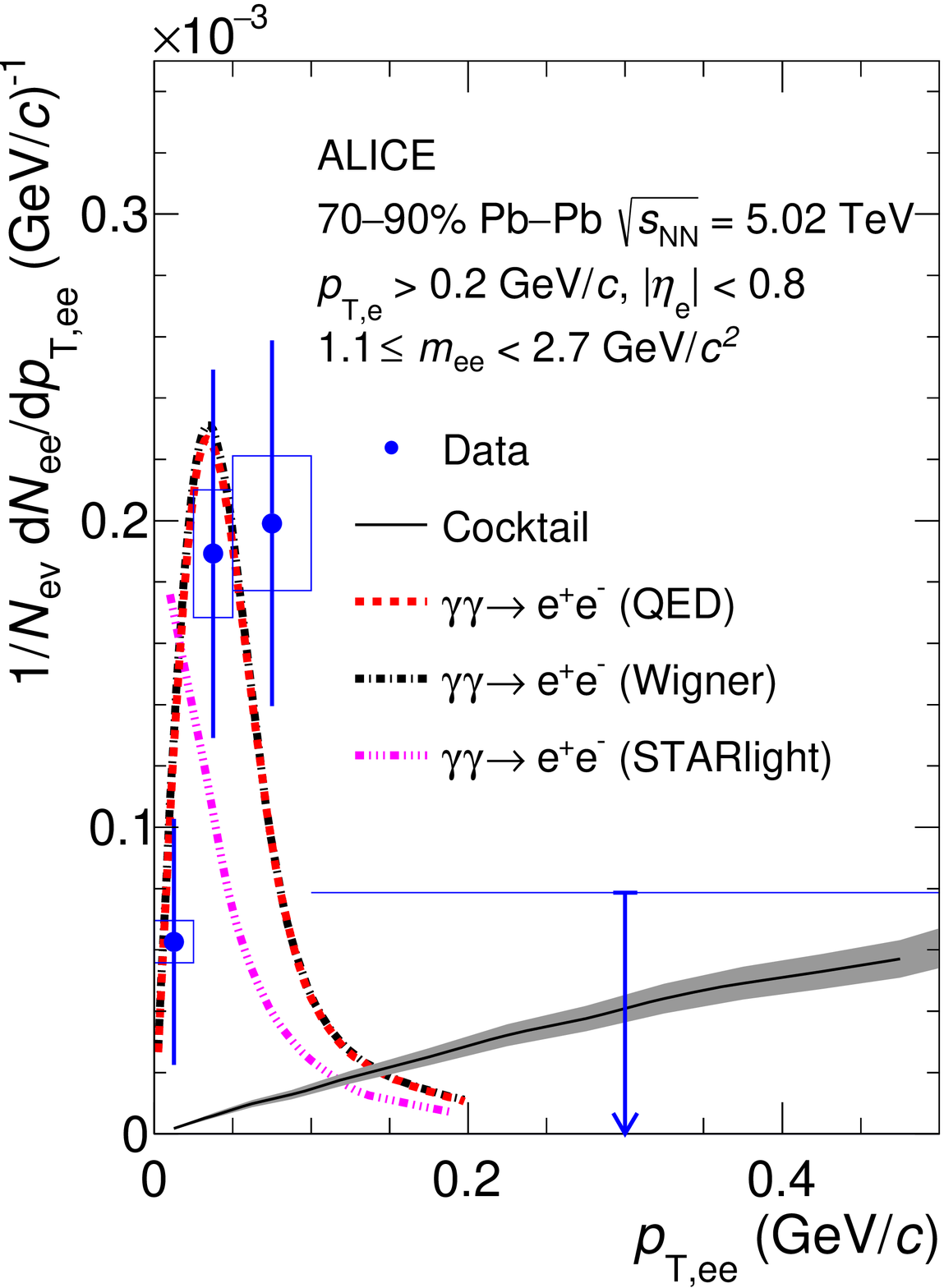}    
    \end{center}
    \caption{Dielectron $p_{\rm T,ee}$-differential yields in peripheral (70--90\%) Pb--Pb collisions at \fivenn for three different \mee ranges, i.e.\ 0.4~$\leq$~\mee~$<$~0.7~\GeVmass (left), 0.7~$\leq$~\mee~$<$~1.1~\GeVmass (middle), and 1.1~$\leq$~\mee~$<$~2.7~\GeVmass (right), compared with the expected \ee contributions from known hadronic decays and calculations for coherent two-photon production of dielectrons folded with the detector resolution\,\cite{Zha:2018tlq,Brandenburg:2021lnj, Klusek-Gawenda:2020eja,Klein:2016yzr,Klein:2018cjh}. For details see the text. The error bars and boxes represent the statistical and systematic uncertainties of the data, respectively, whereas the bands show the uncertainties of the hadronic cocktail. Arrows indicate upper limits at 90\% confidence level.}
    \label{fig:pteewiththeory}
\end{figure}

In order to further investigate the dielectrons produced via photon--photon interactions at low \ptee, the \ptee spectra of inclusive \ee pairs are shown in three different invariant mass ranges in peripheral \PbPb collisions at \fivenn in Fig.~\ref{fig:pteewiththeory}. While the measured yield at $\ptee \geq 0.1$~\GeVc can be described by the hadronic cocktail, a clear peak is seen at \ptee smaller than 0.1~\GeVc in all \mee ranges. The latter is fairly well reproduced by the aforementioned photon--photon models including the impact parameter dependence of the photon $k_{\rm T}$ distribution, i.e.\ the lowest-order QED calculations\,\cite{Zha:2018tlq,Brandenburg:2021lnj} and calculations using the Wigner formalism\,\cite{Klusek-Gawenda:2020eja}. Both approaches predict very similar \ptee distributions. On the contrary, all spectra computed with the STARlight model\,\cite{Klein:2016yzr,Klein:2018cjh} show a rise towards \ptee equal to zero, which is disfavored by the data. By integrating over all transverse distances in the single-photon distribution, the $k_{\rm T}$-factorization approach employed in STARlight leads to a \ptee distribution whose shape is independent of the impact parameter. Such a treatment gives rise to uncertainties on the $k_{\rm T}$ photon distribution of the order of $\omega_{\gamma}$/$\gamma_{\rm L}$, which is precisely the same order of magnitude as $k_{\rm T}$ itself\,\cite{Vidovic:1992ik,Hencken:1994my}. Therefore the $b$ dependence of $k_{\rm T}$, and as a consequence of $p_{\rm T,ee}$, needs to be taken into account in the calculations in order to interpret the results correctly. The limited \pt resolution of the detector has a negligible effect compared to the data uncertainties at low \mee ($0.4 \leq \mee < 0.7$~\GeVmass) but it affects more significantly the reconstructed \ptee distributions at large \mee ($1.1 \leq \mee < 2.7$~\GeVmass). At large \mee, where electrons have larger \pt, the detector resolution on \pt worsens. The reconstructed \ptee distributions are pushed towards larger \ptee values compared to the true \ptee spectra. The maximum of the spectra predicted with the Wigner formalism and lowest-order QED calculations is reduced by about 35\%.

\begin{figure}[tbh]
    \begin{center}
    \includegraphics[width = 0.49\textwidth]{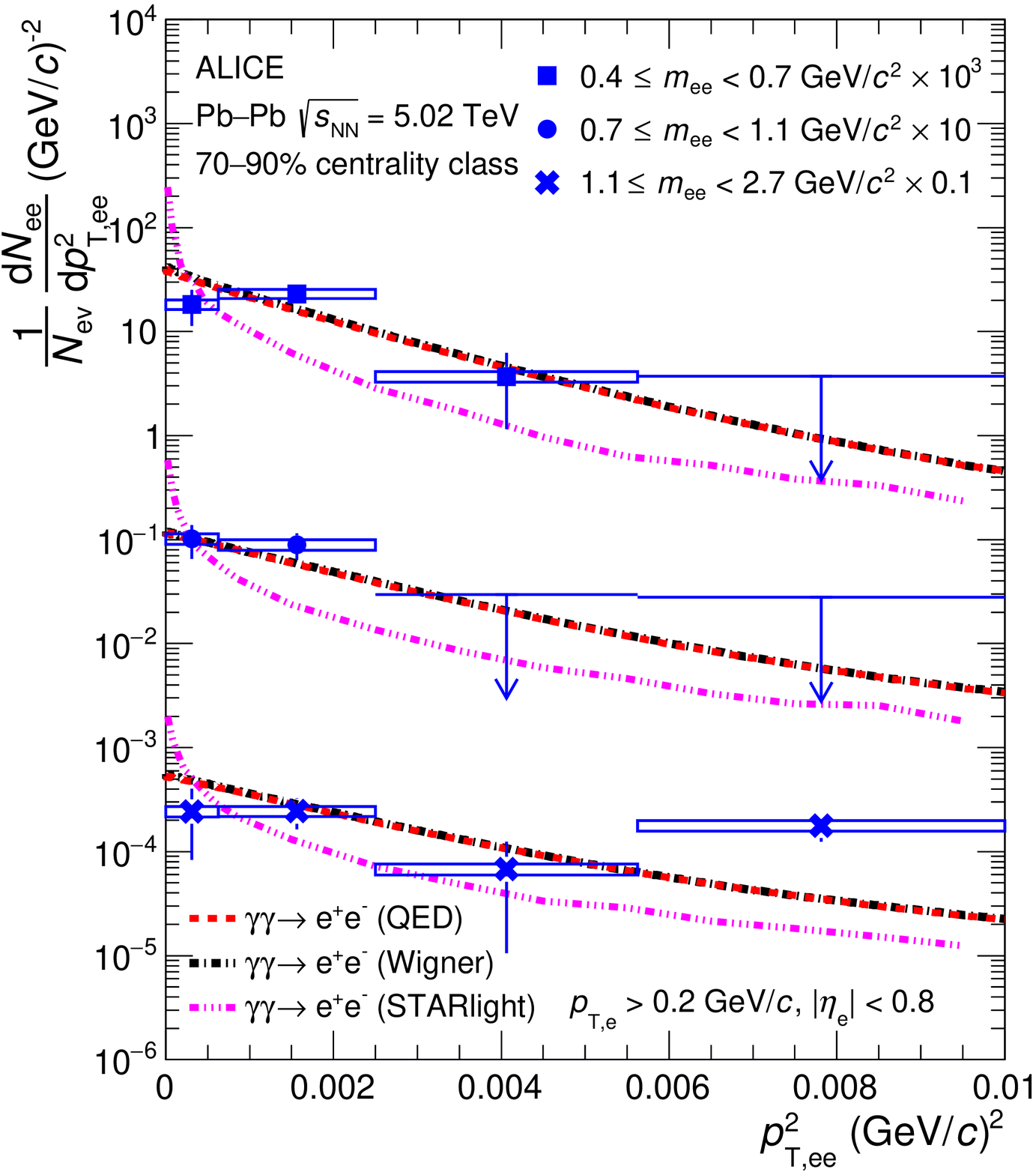}
    \includegraphics[width = 0.49\textwidth]{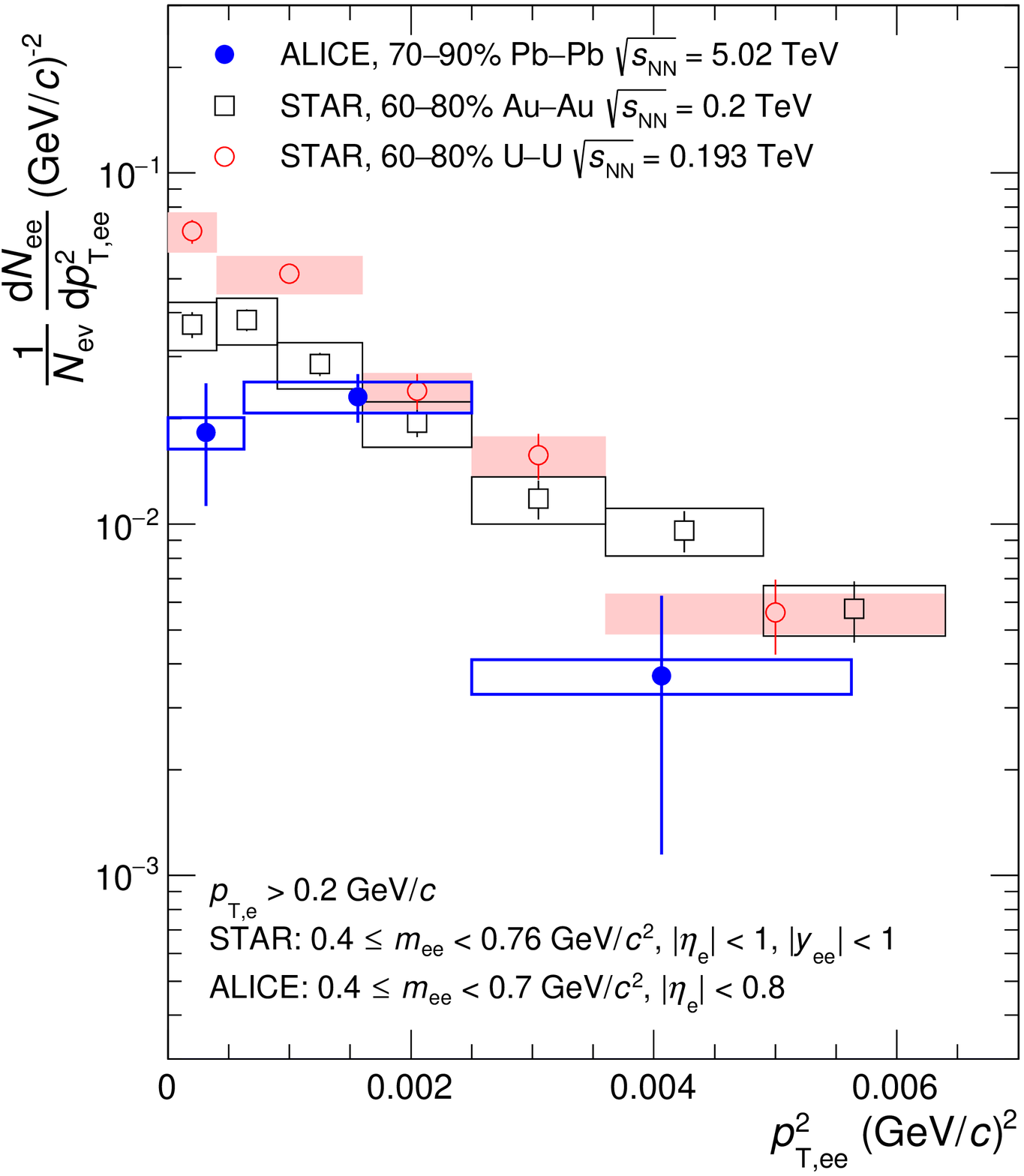}      
    \end{center}
    \caption{Left: Excess dielectron $p_{\rm T,ee}^{\rm 2}$-differential yields after subtraction of the cocktail of known hadronic decay contributions in peripheral (70--90\%) Pb--Pb collisions at \fivenn for different \mee ranges, i.e.\ 0.4~$\leq$~\mee~$<$~0.7~\GeVmass, 0.7~$\leq$~\mee~$<$~1.1~\GeVmass and 1.1~$\leq$~\mee~$<$~2.7~\GeVmass, compared with calculations for coherent photon--photon production of dielectrons folded with the detector resolution\,\cite{Zha:2018tlq,Brandenburg:2021lnj,Klusek-Gawenda:2020eja,Klein:2016yzr,Klein:2018cjh}. Right: Excess dielectron $p_{\rm T,ee}^{\rm 2}$-differential yields after subtraction of the cocktail of known hadronic decay contributions in peripheral Pb--Pb (70--90\%), Au--Au (60--80\%) and U--U (60--80\%) collisions at $\sqrt{s_{\rm NN}}$ $=$ 5.02, 0.2 and 0.193\,TeV\,\cite{STAR:2018ldd}, respectively, in a similar \mee range. The error bars and boxes represent the statistical and systematic uncertainties of the data, respectively. Arrows indicate upper limits at 90\% confidence level.}
    \label{fig:ptsquare}
\end{figure}

The \pteesquare distributions of the excess \ee pairs after subtracting the hadronic cocktail are shown in Fig.~\ref{fig:ptsquare} for the three invariant mass regions in peripheral \PbPb collisions at \fivenn together with the different calculations for photon--photon production of dielectrons\,\cite{Zha:2018tlq,Brandenburg:2021lnj,Klusek-Gawenda:2020eja,Klein:2016yzr,Klein:2018cjh}. The data can be reproduced by the lowest-order QED predictions\,\cite{Zha:2018tlq,Brandenburg:2021lnj} and computations from the authors of Ref.\,\cite{Klusek-Gawenda:2020eja}, whereas the STARlight calculation\,\cite{Klein:2016yzr,Klein:2018cjh} falls below the data points for \pteesquare larger than $6.25 \times 10^{-4}$~(GeV/$c$)$^{2}$ and overshoots the measured spectra at low \pteesquare. This observation is consistent with the results shown  as a function of \ptee and is in line with previous experimental measurements\,\cite{STAR:2019wlg,CMS:2020skx,ATLAS:2020epq} which have demonstrated that the photon $k_{\rm T}$-factorization approach used in STARlight lacks $b$ dependences clearly visible in the experimental measurements. The data support the statement that the \ptee broadening observed in HHICs in comparison to those in UPCs originates predominantly from the initial EM field strength that varies significantly with impact parameter. To quantify the spread of the \ptee distributions, the $\sqrt{\langle \ptee^{2} \rangle}$ is calculated for both the data and aforementioned photon--photon models in the measured \pteesquare range ($0 \leq \pteesquare < 0.01$~(GeV/$c$)$^{2}$). The values are given in Table~\ref{tablepteesquare}. The measured $\sqrt{\langle \ptee^{2} \rangle}$ are found to be in agreement with expectations from theory within uncertainties. The lowest-order QED calculations and the predictions based on the Wigner formalism predict similar $\sqrt{\langle \ptee^{2} \rangle}$ for the three different \mee bins. The increase observed in Table~\ref{tablepteesquare} is mostly due to detector \pt resolution effects. The data are not yet precise enough to conclude on a possible \mee dependence of $\sqrt{\langle \ptee^{2} \rangle}$.

\begin{table}[ht!]
\begin{center}
\centering
\caption{The measured $\sqrt{\langle \ptee^{2} \rangle}$ of excess yields in 70-90\% peripheral \PbPb collisions at \fivenn compared with expectations from photon--photon calculations\,\cite{Zha:2018tlq,Brandenburg:2021lnj,Klusek-Gawenda:2020eja,Klein:2016yzr,Klein:2018cjh}. For details see text.}
\begin{tabular}{lllll}
 \hline 
  Mass region   & Data  &  QED &  Wigner & STARlight \\
   (\GeVmass)   &   & \cite{Zha:2018tlq,Brandenburg:2021lnj} & \cite{Klusek-Gawenda:2020eja}  &  \cite{Klein:2016yzr,Klein:2018cjh} \\
\hline
0.4 $\leq$ \mee $\leq$ 0.7 & 44 $\pm$ 28 (stat.) $\pm$ 6 (syst.)~MeV/$c$  & 44~MeV/$c$ & 45~MeV/$c$ & 30~MeV/$c$\\\hline
0.7 $\leq$ \mee $\leq$ 1.1  & 45 $\pm$ 36 (stat.) $\pm$ 8 (syst.)~MeV/$c$ & 48~MeV/$c$ & 48~MeV/$c$ & 38~MeV/$c$ \\\hline
1.1 $\leq$ \mee $\leq$ 2.7  & 69 $\pm$ 36 (stat.) $\pm$ 8 (syst.)~MeV/$c$ & 50~MeV/$c$  & 50~MeV/$c$ & 42~MeV/$c$\\\hline                                                                
\hline 
\end{tabular}
\label{tablepteesquare}
\end{center}
\end{table}

On the right panel of Fig.~\ref{fig:ptsquare}, the measured \pteesquare spectrum  for $0.4 \leq \mee < 0.7$~\GeVmass in peripheral \PbPb collisions is compared to the \pteesquare distributions measured by the STAR collaboration in a similar phase-space region in peripheral (60-80\%) Au--Au and U--U collisions at $\sqrt{s_{\rm NN}}$ $=$ 200~GeV and 193~GeV\,\cite{STAR:2018ldd}. On the one hand, the \snn dependence of the cross section for the reaction $\gamma\gamma\to{\rm e^{+}}{\rm e^{-}}$ is expected to be rather small from RHIC to LHC energies in the low \mee range and midrapidity region considered here\,\cite{Klusek-Gawenda:2018zfz}. On the other hand, the $Z$ of the different colliding ions are different ($Z_{\rm Au}$ $=$ 79, $Z_{\rm Pb}$ $=$ 82, $Z_{\rm U}$ $=$ 92) and the $\eta_{\rm e}$, $y_{\rm ee}$, and \mee ranges used in the STAR and ALICE experiments are not exactly the same. The results at LHC are found to be similar to the ones at RHIC within large uncertainties. The measured $\sqrt{\langle \ptee^{2} \rangle}$ (see Table~\ref{tablepteesquare}) is comparable to the ones observed in peripheral Au--Au (50.8 $\pm$ 2.51 (stat.+syst.)~MeV/$c$) and U--U (43 $\pm$ 2.26 (stat.+syst.)~MeV/$c$) collisions.

\section{Summary and outlook}
\label{summary}

The first measurements of \ee pairs at low \ptee ($\ptee < 0.1$~\GeVc) and \mee ($0.4 \leq \mee < 2.7$~\GeVmass) at LHC energies are presented at midrapidity (\etarangee{0.8}) in peripheral (70--90\%) and semi-peripheral (50--70\%) \PbPb collisions at $\sqrt{s_{\rm NN}} =$ 5.02\,TeV. An excess of dielectrons is observed at low \ptee over the full measured \mee range compared to the expected \ee yield from known hadronic sources and thermal radiation from the medium in \PbPb collisions. The excess yields after subtraction of the hadronic cocktail do not exhibit a significant centrality dependence and can be reproduced as a function of \mee by different calculations for photon--photon production of dielectrons in both centrality classes.
In peripheral \PbPb collisions the inclusive \ptee spectra and the excess dielectron \pteesquare distributions are shown in three different \mee intervals ($0.4 \leq \mee < 0.7$~\GeVmass, $0.7 \leq \mee < 1.1$~\GeVmass, and $1.1 \leq \mee < 2.7$~\GeVmass) and compared with the hadronic cocktail and predictions for the $\gamma\gamma\to{\rm e^{+}}{\rm e^{-}}$ process using the same models as for the \mee spectra. The results at $\ptee < 0.1$~\GeV/c ($\pteesquare < 0.01$~(\GeVc)$^{2}$) clearly disfavor the shape of the spectra of photon--photon produced dielectrons computed with STARlight\,\cite{Klein:2016yzr,Klein:2018cjh}, whereas they are reproduced by lowest-order QED calculations\,\cite{Zha:2018tlq,Brandenburg:2021lnj} and calculations using the Wigner formalism\,\cite{Klusek-Gawenda:2020eja}. STARlight does not contain any impact-parameter effects on the shape of the transverse momentum distribution of the quasi-real photons and thus on the one of the \ptee and \pteesquare distributions of the produced \ee pairs. According to the calculations\,\cite{Klein:2016yzr,Klein:2018cjh,Zha:2018tlq,Brandenburg:2021lnj,Klusek-Gawenda:2020eja}, these impact-parameter dependencies cannot be neglected in theoretical models computing the $\gamma\gamma\to{\rm l^{+}}{\rm l^{-}}$ process in non ultra-peripheral heavy-ion collisions in order to interpret the data correctly. These results are in line with the statement that the \ptee broadening observed in HHICs in comparison to those in UPCs originates predominantly from the initial electromagnetic field strength that varies significantly with impact parameter. Therefore, determining precisely the magnitude of possible final-state effects related to the creation of a hot and dense medium in HHICs requires a very good understanding of the electromagnetic field produced in heavy-ion collisions. Finally, the measured $\sqrt{\langle \ptee^{2} \rangle}$ in $0.4 \leq \mee < 0.7$~\GeVmass is compatible with the values observed in non-central Au--Au and U--U collisions by STAR at RHIC\,\cite{STAR:2018ldd}.      

A significant improvement in the measurement, as well as more differential studies, are expected after the ALICE upgrades for the LHC Runs 3 and 4, where the number of recorded collisions for the centrality classes considered in this article is expected to increase by a factor greater than 50\,\cite{Musa:1475244,CERN-LHCC-2013-020,CERN-LHCC-2015-002}. The reduced material budget in front of the first tracking layer, together with the improved resolution of the distance-of-closest approach to the collision vertex, will help to suppress the combinatorial and heavy-flavour backgrounds, relevant in such analyses.



\newenvironment{acknowledgement}{\relax}{\relax}
\begin{acknowledgement}
\section*{Acknowledgements}

The ALICE Collaboration would like to thank all its engineers and technicians for their invaluable contributions to the construction of the experiment and the CERN accelerator teams for the outstanding performance of the LHC complex.
The ALICE Collaboration gratefully acknowledges the resources and support provided by all Grid centres and the Worldwide LHC Computing Grid (WLCG) collaboration.
The ALICE Collaboration acknowledges the following funding agencies for their support in building and running the ALICE detector:
A. I. Alikhanyan National Science Laboratory (Yerevan Physics Institute) Foundation (ANSL), State Committee of Science and World Federation of Scientists (WFS), Armenia;
Austrian Academy of Sciences, Austrian Science Fund (FWF): [M 2467-N36] and Nationalstiftung f\"{u}r Forschung, Technologie und Entwicklung, Austria;
Ministry of Communications and High Technologies, National Nuclear Research Center, Azerbaijan;
Conselho Nacional de Desenvolvimento Cient\'{\i}fico e Tecnol\'{o}gico (CNPq), Financiadora de Estudos e Projetos (Finep), Funda\c{c}\~{a}o de Amparo \`{a} Pesquisa do Estado de S\~{a}o Paulo (FAPESP) and Universidade Federal do Rio Grande do Sul (UFRGS), Brazil;
Bulgarian Ministry of Education and Science, within the National Roadmap for Research Infrastructures 2020¿2027 (object CERN), Bulgaria;
Ministry of Education of China (MOEC) , Ministry of Science \& Technology of China (MSTC) and National Natural Science Foundation of China (NSFC), China;
Ministry of Science and Education and Croatian Science Foundation, Croatia;
Centro de Aplicaciones Tecnol\'{o}gicas y Desarrollo Nuclear (CEADEN), Cubaenerg\'{\i}a, Cuba;
Ministry of Education, Youth and Sports of the Czech Republic, Czech Republic;
The Danish Council for Independent Research | Natural Sciences, the VILLUM FONDEN and Danish National Research Foundation (DNRF), Denmark;
Helsinki Institute of Physics (HIP), Finland;
Commissariat \`{a} l'Energie Atomique (CEA) and Institut National de Physique Nucl\'{e}aire et de Physique des Particules (IN2P3) and Centre National de la Recherche Scientifique (CNRS), France;
Bundesministerium f\"{u}r Bildung und Forschung (BMBF) and GSI Helmholtzzentrum f\"{u}r Schwerionenforschung GmbH, Germany;
General Secretariat for Research and Technology, Ministry of Education, Research and Religions, Greece;
National Research, Development and Innovation Office, Hungary;
Department of Atomic Energy Government of India (DAE), Department of Science and Technology, Government of India (DST), University Grants Commission, Government of India (UGC) and Council of Scientific and Industrial Research (CSIR), India;
National Research and Innovation Agency - BRIN, Indonesia;
Istituto Nazionale di Fisica Nucleare (INFN), Italy;
Japanese Ministry of Education, Culture, Sports, Science and Technology (MEXT) and Japan Society for the Promotion of Science (JSPS) KAKENHI, Japan;
Consejo Nacional de Ciencia (CONACYT) y Tecnolog\'{i}a, through Fondo de Cooperaci\'{o}n Internacional en Ciencia y Tecnolog\'{i}a (FONCICYT) and Direcci\'{o}n General de Asuntos del Personal Academico (DGAPA), Mexico;
Nederlandse Organisatie voor Wetenschappelijk Onderzoek (NWO), Netherlands;
The Research Council of Norway, Norway;
Commission on Science and Technology for Sustainable Development in the South (COMSATS), Pakistan;
Pontificia Universidad Cat\'{o}lica del Per\'{u}, Peru;
Ministry of Education and Science, National Science Centre and WUT ID-UB, Poland;
Korea Institute of Science and Technology Information and National Research Foundation of Korea (NRF), Republic of Korea;
Ministry of Education and Scientific Research, Institute of Atomic Physics, Ministry of Research and Innovation and Institute of Atomic Physics and University Politehnica of Bucharest, Romania;
Ministry of Education, Science, Research and Sport of the Slovak Republic, Slovakia;
National Research Foundation of South Africa, South Africa;
Swedish Research Council (VR) and Knut \& Alice Wallenberg Foundation (KAW), Sweden;
European Organization for Nuclear Research, Switzerland;
Suranaree University of Technology (SUT), National Science and Technology Development Agency (NSTDA), Thailand Science Research and Innovation (TSRI) and National Science, Research and Innovation Fund (NSRF), Thailand;
Turkish Energy, Nuclear and Mineral Research Agency (TENMAK), Turkey;
National Academy of  Sciences of Ukraine, Ukraine;
Science and Technology Facilities Council (STFC), United Kingdom;
National Science Foundation of the United States of America (NSF) and United States Department of Energy, Office of Nuclear Physics (DOE NP), United States of America.
In addition, individual groups or members have received support from:
Marie Sk\l{}odowska Curie, Strong 2020 - Horizon 2020, European Research Council (grant nos. 824093, 896850, 950692), European Union;
Academy of Finland (Center of Excellence in Quark Matter) (grant nos. 346327, 346328), Finland;
Programa de Apoyos para la Superaci\'{o}n del Personal Acad\'{e}mico, UNAM, Mexico.

\end{acknowledgement}

\bibliographystyle{utphys}   
\bibliography{bibliography}

\newpage
\appendix

%
%

\section{The ALICE Collaboration}
\label{app:collab}
\begin{flushleft} 
\small

S.~Acharya\,\orcidlink{0000-0002-9213-5329}\,$^{\rm 124,131}$, 
D.~Adamov\'{a}\,\orcidlink{0000-0002-0504-7428}\,$^{\rm 86}$, 
A.~Adler$^{\rm 69}$, 
G.~Aglieri Rinella\,\orcidlink{0000-0002-9611-3696}\,$^{\rm 32}$, 
M.~Agnello\,\orcidlink{0000-0002-0760-5075}\,$^{\rm 29}$, 
N.~Agrawal\,\orcidlink{0000-0003-0348-9836}\,$^{\rm 50}$, 
Z.~Ahammed\,\orcidlink{0000-0001-5241-7412}\,$^{\rm 131}$, 
S.~Ahmad\,\orcidlink{0000-0003-0497-5705}\,$^{\rm 15}$, 
S.U.~Ahn\,\orcidlink{0000-0001-8847-489X}\,$^{\rm 70}$, 
I.~Ahuja\,\orcidlink{0000-0002-4417-1392}\,$^{\rm 37}$
% ...

\section*{Affiliation Notes}

$^{\rm I}$ Deceased\\
$^{\rm II}$ Also at: Max-Planck-Institut f\"{u}r Physik, Munich, Germany\\
$^{\rm III}$ Also at: Italian National Agency for New Technologies, Energy and Sustainable Economic Development (ENEA), Bologna, Italy\\
$^{\rm IV}$ Also at: Dipartimento DET del Politecnico di Torino, Turin, Italy\\
$^{\rm V}$ Also at: Department of Applied Physics, Aligarh Muslim University, Aligarh, India\\
$^{\rm VI}$ Also at: Institute of Theoretical Physics, University of Wroclaw, Poland\\
$^{\rm VII}$ Also at: An institution covered by a cooperation agreement with CERN\\

\section*{Collaboration Institutes}

$^{1}$ A.I. Alikhanyan National Science Laboratory (Yerevan Physics Institute) Foundation, Yerevan, Armenia\\
$^{2}$ AGH University of Science and Technology, Cracow, Poland\\
$^{3}$ Bogolyubov Institute for Theoretical Physics, National Academy of Sciences of Ukraine, Kiev, Ukraine\\
$^{4}$ Bose Institute, Department of Physics  and Centre for Astroparticle Physics and Space Science (CAPSS), Kolkata, India\\
$^{5}$ California Polytechnic State University, San Luis Obispo, California, United States\\
$^{6}$ Central China Normal University, Wuhan, China\\
$^{7}$ Centro de Aplicaciones Tecnol\'{o}gicas y Desarrollo Nuclear (CEADEN), Havana, Cuba\\
$^{8}$ Centro de Investigaci\'{o}n y de Estudios Avanzados (CINVESTAV), Mexico City and M\'{e}rida, Mexico\\
$^{9}$ Chicago State University, Chicago, Illinois, United States\\
$^{10}$ China Institute of Atomic Energy, Beijing, China\\
$^{11}$ Chungbuk National University, Cheongju, Republic of Korea\\
$^{12}$ Comenius University Bratislava, Faculty of Mathematics, Physics and Informatics, Bratislava, Slovak Republic\\
$^{13}$ COMSATS University Islamabad, Islamabad, Pakistan\\
$^{14}$ Creighton University, Omaha, Nebraska, United States\\
$^{15}$ Department of Physics, Aligarh Muslim University, Aligarh, India\\
$^{16}$ Department of Physics, Pusan National University, Pusan, Republic of Korea\\
$^{17}$ Department of Physics, Sejong University, Seoul, Republic of Korea\\
$^{18}$ Department of Physics, University of California, Berkeley, California, United States\\
$^{19}$ Department of Physics, University of Oslo, Oslo, Norway\\
$^{20}$ Department of Physics and Technology, University of Bergen, Bergen, Norway\\
$^{21}$ Dipartimento di Fisica, Universit\`{a} di Pavia, Pavia, Italy\\
$^{22}$ Dipartimento di Fisica dell'Universit\`{a} and Sezione INFN, Cagliari, Italy\\
$^{23}$ Dipartimento di Fisica dell'Universit\`{a} and Sezione INFN, Trieste, Italy\\
$^{24}$ Dipartimento di Fisica dell'Universit\`{a} and Sezione INFN, Turin, Italy\\
$^{25}$ Dipartimento di Fisica e Astronomia dell'Universit\`{a} and Sezione INFN, Bologna, Italy\\
$^{26}$ Dipartimento di Fisica e Astronomia dell'Universit\`{a} and Sezione INFN, Catania, Italy\\
$^{27}$ Dipartimento di Fisica e Astronomia dell'Universit\`{a} and Sezione INFN, Padova, Italy\\
$^{28}$ Dipartimento di Fisica `E.R.~Caianiello' dell'Universit\`{a} and Gruppo Collegato INFN, Salerno, Italy\\
$^{29}$ Dipartimento DISAT del Politecnico and Sezione INFN, Turin, Italy\\
$^{30}$ Dipartimento di Scienze MIFT, Universit\`{a} di Messina, Messina, Italy\\
$^{31}$ Dipartimento Interateneo di Fisica `M.~Merlin' and Sezione INFN, Bari, Italy\\
$^{32}$ European Organization for Nuclear Research (CERN), Geneva, Switzerland\\
$^{33}$ Faculty of Electrical Engineering, Mechanical Engineering and Naval Architecture, University of Split, Split, Croatia\\
$^{34}$ Faculty of Engineering and Science, Western Norway University of Applied Sciences, Bergen, Norway\\
$^{35}$ Faculty of Nuclear Sciences and Physical Engineering, Czech Technical University in Prague, Prague, Czech Republic\\
$^{36}$ Faculty of Physics, Sofia University, Sofia, Bulgaria\\
$^{37}$ Faculty of Science, P.J.~\v{S}af\'{a}rik University, Ko\v{s}ice, Slovak Republic\\
$^{38}$ Frankfurt Institute for Advanced Studies, Johann Wolfgang Goethe-Universit\"{a}t Frankfurt, Frankfurt, Germany\\
$^{39}$ Fudan University, Shanghai, China\\
$^{40}$ Gangneung-Wonju National University, Gangneung, Republic of Korea\\
$^{41}$ Gauhati University, Department of Physics, Guwahati, India\\
$^{42}$ Helmholtz-Institut f\"{u}r Strahlen- und Kernphysik, Rheinische Friedrich-Wilhelms-Universit\"{a}t Bonn, Bonn, Germany\\
$^{43}$ Helsinki Institute of Physics (HIP), Helsinki, Finland\\
$^{44}$ High Energy Physics Group,  Universidad Aut\'{o}noma de Puebla, Puebla, Mexico\\
$^{45}$ Horia Hulubei National Institute of Physics and Nuclear Engineering, Bucharest, Romania\\
$^{46}$ Indian Institute of Technology Bombay (IIT), Mumbai, India\\
$^{47}$ Indian Institute of Technology Indore, Indore, India\\
$^{48}$ INFN, Laboratori Nazionali di Frascati, Frascati, Italy\\
$^{49}$ INFN, Sezione di Bari, Bari, Italy\\
$^{50}$ INFN, Sezione di Bologna, Bologna, Italy\\
$^{51}$ INFN, Sezione di Cagliari, Cagliari, Italy\\
$^{52}$ INFN, Sezione di Catania, Catania, Italy\\
$^{53}$ INFN, Sezione di Padova, Padova, Italy\\
$^{54}$ INFN, Sezione di Pavia, Pavia, Italy\\
$^{55}$ INFN, Sezione di Torino, Turin, Italy\\
$^{56}$ INFN, Sezione di Trieste, Trieste, Italy\\
$^{57}$ Inha University, Incheon, Republic of Korea\\
$^{58}$ Institute for Gravitational and Subatomic Physics (GRASP), Utrecht University/Nikhef, Utrecht, Netherlands\\
$^{59}$ Institute of Experimental Physics, Slovak Academy of Sciences, Ko\v{s}ice, Slovak Republic\\
$^{60}$ Institute of Physics, Homi Bhabha National Institute, Bhubaneswar, India\\
$^{61}$ Institute of Physics of the Czech Academy of Sciences, Prague, Czech Republic\\
$^{62}$ Institute of Space Science (ISS), Bucharest, Romania\\
$^{63}$ Institut f\"{u}r Kernphysik, Johann Wolfgang Goethe-Universit\"{a}t Frankfurt, Frankfurt, Germany\\
$^{64}$ Instituto de Ciencias Nucleares, Universidad Nacional Aut\'{o}noma de M\'{e}xico, Mexico City, Mexico\\
$^{65}$ Instituto de F\'{i}sica, Universidade Federal do Rio Grande do Sul (UFRGS), Porto Alegre, Brazil\\
$^{66}$ Instituto de F\'{\i}sica, Universidad Nacional Aut\'{o}noma de M\'{e}xico, Mexico City, Mexico\\
$^{67}$ iThemba LABS, National Research Foundation, Somerset West, South Africa\\
$^{68}$ Jeonbuk National University, Jeonju, Republic of Korea\\
$^{69}$ Johann-Wolfgang-Goethe Universit\"{a}t Frankfurt Institut f\"{u}r Informatik, Fachbereich Informatik und Mathematik, Frankfurt, Germany\\
$^{70}$ Korea Institute of Science and Technology Information, Daejeon, Republic of Korea\\
$^{71}$ KTO Karatay University, Konya, Turkey\\
$^{72}$ Laboratoire de Physique des 2 Infinis, Ir\`{e}ne Joliot-Curie, Orsay, France\\
$^{73}$ Laboratoire de Physique Subatomique et de Cosmologie, Universit\'{e} Grenoble-Alpes, CNRS-IN2P3, Grenoble, France\\
$^{74}$ Lawrence Berkeley National Laboratory, Berkeley, California, United States\\
$^{75}$ Lund University Department of Physics, Division of Particle Physics, Lund, Sweden\\
$^{76}$ Nagasaki Institute of Applied Science, Nagasaki, Japan\\
$^{77}$ Nara Women{'}s University (NWU), Nara, Japan\\
$^{78}$ National and Kapodistrian University of Athens, School of Science, Department of Physics , Athens, Greece\\
$^{79}$ National Centre for Nuclear Research, Warsaw, Poland\\
$^{80}$ National Institute of Science Education and Research, Homi Bhabha National Institute, Jatni, India\\
$^{81}$ National Nuclear Research Center, Baku, Azerbaijan\\
$^{82}$ National Research and Innovation Agency - BRIN, Jakarta, Indonesia\\
$^{83}$ Niels Bohr Institute, University of Copenhagen, Copenhagen, Denmark\\
$^{84}$ Nikhef, National institute for subatomic physics, Amsterdam, Netherlands\\
$^{85}$ Nuclear Physics Group, STFC Daresbury Laboratory, Daresbury, United Kingdom\\
$^{86}$ Nuclear Physics Institute of the Czech Academy of Sciences, Husinec-\v{R}e\v{z}, Czech Republic\\
$^{87}$ Oak Ridge National Laboratory, Oak Ridge, Tennessee, United States\\
$^{88}$ Ohio State University, Columbus, Ohio, United States\\
$^{89}$ Physics department, Faculty of science, University of Zagreb, Zagreb, Croatia\\
$^{90}$ Physics Department, Panjab University, Chandigarh, India\\
$^{91}$ Physics Department, University of Jammu, Jammu, India\\
$^{92}$ Physics Department, University of Rajasthan, Jaipur, India\\
$^{93}$ Physics Program and International Institute for Sustainability with Knotted Chiral Meta Matter (SKCM2), Hiroshima University, Hiroshima, Japan\\
$^{94}$ Physikalisches Institut, Eberhard-Karls-Universit\"{a}t T\"{u}bingen, T\"{u}bingen, Germany\\
$^{95}$ Physikalisches Institut, Ruprecht-Karls-Universit\"{a}t Heidelberg, Heidelberg, Germany\\
$^{96}$ Physik Department, Technische Universit\"{a}t M\"{u}nchen, Munich, Germany\\
$^{97}$ Politecnico di Bari and Sezione INFN, Bari, Italy\\
$^{98}$ Research Division and ExtreMe Matter Institute EMMI, GSI Helmholtzzentrum f\"ur Schwerionenforschung GmbH, Darmstadt, Germany\\
$^{99}$ Saha Institute of Nuclear Physics, Homi Bhabha National Institute, Kolkata, India\\
$^{100}$ School of Physics and Astronomy, University of Birmingham, Birmingham, United Kingdom\\
$^{101}$ Secci\'{o}n F\'{\i}sica, Departamento de Ciencias, Pontificia Universidad Cat\'{o}lica del Per\'{u}, Lima, Peru\\
$^{102}$ Stefan Meyer Institut f\"{u}r Subatomare Physik (SMI), Vienna, Austria\\
$^{103}$ SUBATECH, IMT Atlantique, Nantes Universit\'{e}, CNRS-IN2P3, Nantes, France\\
$^{104}$ Suranaree University of Technology, Nakhon Ratchasima, Thailand\\
$^{105}$ Technical University of Ko\v{s}ice, Ko\v{s}ice, Slovak Republic\\
$^{106}$ The Henryk Niewodniczanski Institute of Nuclear Physics, Polish Academy of Sciences, Cracow, Poland\\
$^{107}$ The University of Texas at Austin, Austin, Texas, United States\\
$^{108}$ Universidad Aut\'{o}noma de Sinaloa, Culiac\'{a}n, Mexico\\
$^{109}$ Universidade de S\~{a}o Paulo (USP), S\~{a}o Paulo, Brazil\\
$^{110}$ Universidade Estadual de Campinas (UNICAMP), Campinas, Brazil\\
$^{111}$ Universidade Federal do ABC, Santo Andre, Brazil\\
$^{112}$ University of Cape Town, Cape Town, South Africa\\
$^{113}$ University of Houston, Houston, Texas, United States\\
$^{114}$ University of Jyv\"{a}skyl\"{a}, Jyv\"{a}skyl\"{a}, Finland\\
$^{115}$ University of Kansas, Lawrence, Kansas, United States\\
$^{116}$ University of Liverpool, Liverpool, United Kingdom\\
$^{117}$ University of Science and Technology of China, Hefei, China\\
$^{118}$ University of South-Eastern Norway, Kongsberg, Norway\\
$^{119}$ University of Tennessee, Knoxville, Tennessee, United States\\
$^{120}$ University of the Witwatersrand, Johannesburg, South Africa\\
$^{121}$ University of Tokyo, Tokyo, Japan\\
$^{122}$ University of Tsukuba, Tsukuba, Japan\\
$^{123}$ University Politehnica of Bucharest, Bucharest, Romania\\
$^{124}$ Universit\'{e} Clermont Auvergne, CNRS/IN2P3, LPC, Clermont-Ferrand, France\\
$^{125}$ Universit\'{e} de Lyon, CNRS/IN2P3, Institut de Physique des 2 Infinis de Lyon, Lyon, France\\
$^{126}$ Universit\'{e} de Strasbourg, CNRS, IPHC UMR 7178, F-67000 Strasbourg, France, Strasbourg, France\\
$^{127}$ Universit\'{e} Paris-Saclay Centre d'Etudes de Saclay (CEA), IRFU, D\'{e}partment de Physique Nucl\'{e}aire (DPhN), Saclay, France\\
$^{128}$ Universit\`{a} degli Studi di Foggia, Foggia, Italy\\
$^{129}$ Universit\`{a} del Piemonte Orientale, Vercelli, Italy\\
$^{130}$ Universit\`{a} di Brescia, Brescia, Italy\\
$^{131}$ Variable Energy Cyclotron Centre, Homi Bhabha National Institute, Kolkata, India\\
$^{132}$ Warsaw University of Technology, Warsaw, Poland\\
$^{133}$ Wayne State University, Detroit, Michigan, United States\\
$^{134}$ Westf\"{a}lische Wilhelms-Universit\"{a}t M\"{u}nster, Institut f\"{u}r Kernphysik, M\"{u}nster, Germany\\
$^{135}$ Wigner Research Centre for Physics, Budapest, Hungary\\
$^{136}$ Yale University, New Haven, Connecticut, United States\\
$^{137}$ Yonsei University, Seoul, Republic of Korea\\
$^{138}$  Zentrum  f\"{u}r Technologie und Transfer (ZTT), Worms, Germany\\
$^{139}$ Affiliated with an institute covered by a cooperation agreement with CERN\\
$^{140}$ Affiliated with an international laboratory covered by a cooperation agreement with CERN.\\

\end{flushleft} 

\end{document}